# Substitutional effects in TiFe for hydrogen storage: a comprehensive review

Erika M. Dematteis,[a)b)] Nicola Berti,[b)] Fermin Cuevas,[a)*] Michel Latroche,[a)] and Marcello Baricco[b)]

[a)] Univ Paris Est Creteil, CNRS, ICMPE, 2 rue Henri Dunant, 94320 Thiais, France

[b)] Department of Chemistry, Inter-departmental Center Nanostructured Interfaces and Surfaces (NIS), and INSTM, University of Turin, Via Pietro Giuria 7, 10125 Torino, Italy

*Corresponding author

Fermin Cuevas

E-mail address: cuevas@icmpe.cnrs.fr

Tel.: +33 149781225

**Abstract**

The search for suitable materials for solid-state stationary storage of green hydrogen is pushing the implementation of efficient renewable energy systems. This involves rational design and modification of cheap alloys for effective storage in mild conditions of temperature and pressure. Among many intermetallic compounds described in the literature, TiFe-based systems have recently regained vivid interest as materials for practical applications since they are low-cost and they can be tuned to match required pressure and operation conditions. This work aims to provide a comprehensive review of publications involving chemical substitution in TiFe-based compounds for guiding compound design and materials selection in current and future hydrogen storage applications. Mono- and multi-substituted compounds modify TiFe thermodynamics and are beneficial for many hydrogenation properties. They will be reviewed and deeply discussed, with a focus on manganese substitution.

**Keywords**: TiFe, titanium, iron, substitution, intermetallic compound, hydrogen storage, metal hydride

## Summary

## Introduction

The hydrogen energy chain is foreseen as one of the key technologies to face the issues of climate change and scarce oil resources. Hydrogen can be worldwide and cleanly produced through electrolysis of water using renewable primary energies. If not consumed on-site, it can be transported by gas pipelines, trucks and ships. Finally, hydrogen can be used to feed fuel cells and generate electricity (and heat) on demand, releasing only water as a by-product and then closing the hydrogen cycle. Such an electricity-hydrogen-electricity conversion process is only sustainable if electricity is produced from renewable energies and cannot be directly injected in the grid, then it can be used later on with a fuel cell. Therefore, it is mandatory to add a storage step in the hydrogen chain for the time-management of hydrogen production and use. The intrinsic intermittency of most renewable energy sources makes unavoidable the implementation of efficient hydrogen storage systems.

Efficient hydrogen storage can be achieved as dihydrogen molecules in high-pressure tanks (typically 350 or 700 bar) or in liquid state (at temperature lower than -252 °C [21 K]).[1–3] In addition, dihydrogen can be physically adsorbed in high-surface-area solids such as MOFs and activated carbons, typically at liquid nitrogen temperature (-196 °C [77 K]).[4–8] As an alternative, hydrogen molecules can be chemisorbed at the surface of solid compounds and diffused in atomic form to form hydrogen-containing compounds.[9–11] In some cases, after suitable activation, these compounds can reversibly absorb and desorb hydrogen close to normal conditions of pressure and temperature (*i.e.* 1 bar, 25 °C).[12–16] These materials are typically intermetallic compounds of general formula $AB_n$, which are commonly named metallic hydrides, from the facts that both hydrogen-metal bonding and electronic conductivity have a metallic character.[17] *A* is an element that forms very stable metallic hydrides (e.g. rare earths and early transition metals) and *B* an element that only forms hydrides at very high pressure (e.g. late transition metals), as reported in **Figure 1**. Their combination in stoichiometric ratio $n = B/A$ allows for the formation of hydrides with intermediate stability.

Representative intermetallic compounds suitable for hydrogen storage are $LaNi_5$, $CeNi_3$, $TiMn_2$ and TiFe for $n$ = 5, 3, 2 and 1, respectively.

Intermetallic compounds, being formed by heavy elements, offer modest mass storage capacities (*i.e.* 1-2 wt.%). Hydrogen systems based on this technology have low gravimetric capacities, due to the weight of the reservoir and ancillary equipment.[3] When compared to classical molecular methods (5-6 wt.% system basis, for both pressurized and liquid storage), their typical operation conditions (0-80°C, 1-50 bar) guarantee higher safety conditions. This is a key property when hydrogen tanks must be installed close to domestic facilities or in confined space. Moreover, in the case of stationary applications, footprint instead of mass capacity is the most relevant performance indicator for the hydrogen storage system. The volumetric capacity of intermetallic compounds, *i.e.* 100-120 $kg_{H_2}/m^3$, is significantly higher than that of pressurized or liquid hydrogen: 39 (at 700 bar) and 70 $kg_{H_2}/m^3$, respectively. Furthermore, if the required tank is considered, the system volumetric density decreases significantly in the case of gas and liquid storage.

Finally, yet importantly, intermetallic compounds are highly versatile materials as their operation temperature and pressure can be tuned at will, through suitable chemical substitutions of both *A* and *B*-type elements. As an example, the EU-funded HyCARE (Hydrogen CArrier for Renewable Energy storage) project, kicked off in January 2019, aims to develop a prototype large-scale hydrogen storage tank using a solid-state hydrogen carrier based on metal powder, operating at low pressure and temperature.[18,19] The project involves the production of almost 4 tons of metal powder, which will be placed in stainless steel containers. The thermal management of the plant will follow an innovative approach, making use of phase-change materials, significantly increasing the efficiency of the process. The aim is to store about 50 kg of hydrogen, which is a rather high quantity to be stored using this technique.

Among many intermetallic compounds described in the literature for reversible hydrogen storage at room temperature (RT), TiFe-based systems have recently regained vivid interest. With a mass and volumetric capacity for the binary compound of 1.87 wt.% and 105 $kg_{H_2}/m^3$, respectively,[20] the relevance of this system is mainly driven by its low cost, as compared to other intermetallics. However, in 2020, the European Union's (EU) has updated a list of 30 critical raw materials (CRMs), including titanium, considering their supply risk and economical importance.[21] In fact, titanium is widely exploited in aeronautics and medical applications, and its processing is making EU strongly dependent on import from main global producers (45% China, 22% Russia, 22% Japan).[21] Titanium End of Life Recycling Input Rate (EoL-RIR) is still limited (reported as 19%),[21] however, for applications as solid state hydrogen storage material, it could be recycled and reused effectively. **Figure 2** shows the CRMs, highlighted with an orange frame, which should thus be avoided or limited in alloy formulation towards large-scale production. Economic and supply indicators demonstrate that TiFe-based compounds are today target materials for practical applications, as shown by the implementation of R&D projects worldwide, and especially in Japan. Intermetallic TiFe compound is promising for hydrogen storage tanks thank to its high volumetric density, good sorption kinetic, reversibility and because it can work in mild temperature and pressure conditions. Moreover, the hydrogenation thermodynamics of TiFe must be tuned to the required conditions of pressure and temperature imposed by each specific hydrogen storage application. As stated above, this can be achieved through suitable atomic substitutions. Indeed, titanium (Ti) and iron (Fe) can be substituted by other elements within certain homogeneity composition ranges, as reported in ternary phase diagrams.[22] Numerous mono and multi-substituted alloys have been explored in the literature as highlighted in green in **Figure 2**. These substitutions have not only a noticeable effect on hydrogen sorption thermodynamics, but also on other key properties, such as alloy activation, reaction kinetics and cycle life.

Recently, Sujan *et al.* provided a review focused on binary TiFe compound and its hydrogenation properties,[23] while Lys *et al.* reported in a short review the effect of substitution on the hydrogenation properties of $A_xB_y$ alloys.[24] Here, after a short overview of binary TiFe, we focus our attention on reviewing the literature on substitutional effects, which are fundamental for practical applications, aiming at extending the previous reviews work in a comprehensive manner. Mono-substituted compounds are discussed with a focus on manganese substitution, which has been proved to be a key element. Some examples of prominent multi-substituted alloys are also reported here. This work aims to provide a comprehensive analysis of the many publications involving chemical substitution in TiFe-based compounds. As a conclusion, some correlations between compositions and hydrogen sorption properties are drawn, for guiding compound design and selection in current and future hydrogen storage applications.

# TiFe

## Synthesis and crystal structures of TiFe and its hydrides

TiFe exhibits a narrow homogeneity range, with the largest domain extending from 49.7 to 52.5 at.% Ti at the eutectic temperature of 1085 °C.[25,26]. TiFe neighbouring phases are $TiFe_2$, at the Fe-rich side, and β-Ti with a maximum solubility of 21 at.% Fe, at the Ti-rich one. The hydrogen storage properties of this intermetallic compound are strictly linked to the composition and to the presence of secondary phases. In fact, due to the composition range in which TiFe phase can be formed, different properties can be observed for stoichiometric TiFe, Fe-rich ($TiFe_{1.012}$) or Ti-rich ($TiFe_{0.905}$) alloys. In addition, for Ti-rich alloys, the formation of β-Ti precipitates enables the hydrogen sorption at RT without activation. The TiFe heat of formation measured at 1167 °C is $\Delta_f H$ = -31.0 kJ $mol^{-1}$.[27]

TiFe is usually produced by melting the elements in a high temperature furnace.[28] As shown in the titanium-iron phase diagram (**Figure 3**), the binary compound is obtained from the melt through a Liquid + $TiFe_2 \rightarrow$ TiFe peritectic reaction at 1317 °C. As an alternative to melting, TiFe can also be obtained and processed by Severe Plastic Deformation (SPD) techniques, such as ball milling[29–45] and high pressure torsion [46–50], as well as by self-ignition.[51–53]. SPD techniques lead to fresh and defective surfaces, that help alloy activation and also nanostructuration, but reducing nominal capacity.[41,54,55]

The crystal structure of TiFe was first identified as CsCl-type (B2, space group *Pm*-3*m*) by Laves *et al.*[56]. Further studies confirmed a cubic lattice, but contrary to *bcc* alloys, with distinct ordered atoms at the cube vertices, 1*a* sites (0,0,0), and its center, 1*b* sites (½, ½ , ½).[57] The lattice constant of the stoichiometric compound is reported to be 2.9763 Å.[58] As stated before, the phase diagram exhibits a small homogeneity domain showing that the crystal structure of TiFe can accommodate some defects, such as partial substitution of Ti on the Fe sites.[59,60]

Regarding the hydrides, four different phases have been reported for the TiFe-H system: α-solid solution, two monohydrides ($\beta_1$ and $\beta_2$) and γ-dihydride. The solid solution retains the CsCl-type structure with minor changes in the lattice parameters.[61] Neutron diffraction experiments showed that H atoms occupy the octahedral 3*d* sites (½, 0, 0) along the cube edges with $Ti_4Fe_2$ coordination, exhibiting a shorter distance from Fe atoms than Ti atoms (1.49 Å *vs*. 2.11 Å). Even though Ti is known for its stronger hydrogen affinity compared to iron, this feature is common for all phases in the TiFe-H system.[62–65] The maximum solubility of hydrogen in the α-solid solution is $TiFeH_{0.1}$. Both β-phases crystallize in an orthorhombic structure, with minor structural differences between them except for hydrogen content, being TiFeH and $TiFeH_{1.4}$ for $\beta_1$ and $\beta_2$, respectively. The most advanced studies by neutron diffraction suggest a $P222_1$ space group,[66] though $Pmc2_1$ cannot be completely ruled out.[67] In both monohydrides, H atoms partially occupy the octahedral sites H1 and H2, both with coordination $Ti_4Fe_2$, whereas Fe is found at site 2*c* (0, 0.294, ¼) and Ti at site 2*d* (½, 0.757, ¼). In the $\beta_1$-phase, hydrogen shows occupancies of 88% and 12% for sites H1 and H2, respectively, whereas in $\beta_2$-phase, it exhibits occupancies of 92% and 45%, respectively. Due to the difficulties in achieving the fully hydrogenated phase, and the broadening of diffraction peaks because of strains, the crystal structure of the γ-phase has been subject of debate. In a first study of Reilly *et al.*,[68] a cubic structure was suggested. Subsequent analyses proposed a monoclinic structure.[62,63,69,70] Later, Fischer *et al.* [64] found an orthorhombic structure (space group *Cmmm*), which was confirmed by both experimental works and theoretical calculations.[71,72] Here, Ti atoms occupy site 4*h* (0.223, 0, ½), Fe site 4*i* (0, 0.2887, 0), while H atoms occupy three different octahedral sites, two of them fully occupied with $Ti_4Fe_2$ coordination and the third one, with $Ti_2Fe_4$ coordination, partially occupied at 91%.

**Activation**

The main drawback for practical application of TiFe is probably the laborious treatment required after synthesis to promote the first hydrogen absorption. This treatment, usually named “activation”, has been the subject of extensive work and controversy. The fact that TiFe does not readily absorb hydrogen at RT has been attributed to a native passivating layer, which forms at its surface. Indeed, TiFe is sensitive to air moisture and might react with it, forming oxides and hydroxides and then hindering the reaction with hydrogen. Consequently, one needs to apply harsh conditions to induce hydrogen penetration, to break the passivating surface layer or to avoid its formation at the alloy surface.

The first description of an activation process for TiFe was reported in the pioneering work of Reilly *et al.*[68]. The authors performed a sequence of absorption (up to 65 bar) and desorption (under vacuum) cycles at both high (400 – 450 °C) and room temperatures. Upon triggering hydrogen absorption/desorption cycles, TiFe undergoes expansion and contraction, respectively, leading to volume changes and inducing the crack of the passivating layer. Since TiFe is a brittle material, fresh and clean TiFe surfaces are uncovered, where hydrogen can promptly be absorbed.

Subsequently, several authors tried to identify the species formed during the oxidation and after activation, with the purpose of better understanding the mechanisms involved in this treatment. Pande *et al.* investigated the surface of oxidized TiFe by means of electron microscopy.[73] In the electron diffraction patterns, they found a phase claimed to be $Ti_3Fe_3O$, because this oxide was found unable to absorb hydrogen, making it a relevant candidate as passivating layer.[74] Bläsius *et al.* studied the surface of activated TiFe by Mössbauer spectroscopy, revealing the presence of small Fe clusters.[75] Because the signal of iron oxide was not detected, they inferred that only titanium was oxidized. Fe precipitates at the alloy surface were also found by other authors, and this free Fe was claimed to play a catalytic role in the dissociation of hydrogen.[76–79] However, Schober *et al.* studied the activation process by TEM and detected $TiO_2$, $TiFe_2$ and suboxide $TiFeO_x$ as surface species.[80] They did not observe any elemental Fe and concluded that Fe clusters are only formed after severe oxygen contamination, following the $TiFe_2 + O_2 \rightarrow TiO_2 + 2Fe$ reaction. Hiebl *et al.* demonstrated that $Ti_2FeO_x$ can absorb hydrogen[81] and other authors detected this compound during annealing of oxidized TiFe,[82,83] casting doubts on the catalytic effect of Fe. Later on, Schlapbach *et al.* identified an oxide layer mainly consisting of $Fe^{III}$ and $Ti^{<IV}$ on the surface of passivated TiFe.[84] After heating, they noticed the formation of Fe and $TiO_2$, suggesting that $TiO_2$ is not an effective catalyst for the reaction, but rather a support for Fe clusters which might split hydrogen molecules.

Reilly *et al.*[58] showed that the composition of the surface layer strongly depends on the annealing conditions and the quantity of oxygen that could be present in the raw materials, synthesis atmosphere or thermal treatment atmosphere. This observation partially explains the controversies in the literature, where each research group followed different treatments. Nonetheless, the procedure suggested by Reilly *et al.* to ensure TiFe activation remained highly laborious for practical activation. To simplify the activation, Chu *et al.*[85] synthesized TiFe by means of mechanical alloying, starting from elemental powders of Ti and Fe. The authors prepared an equiatomic TiFe mixture, ball-milled for different duration times. They got amorphous materials that required one hour annealing at 300 °C under 7 bar $H_2$ for activation. Hotta *et al.*[29] also produced TiFe by ball milling pure Ti and Fe, which required an activation at 300 °C and 150 bar of $H_2$. However, compared to the work of Chu *et al.*, Hotta *et al.* obtained crystalline TiFe that absorbed ~ 1 wt.% $H_2$. Zaluski *et al.* [32] ball milled Ti and Fe, noticing that the final structure of the composite strongly depends on oxygen contamination. For an oxygen content below 3 at.%, TiFe crystallized in the expected CsCl-type structure, whereas at higher oxygen content it became amorphous. Still, both samples required a high temperature to get activated: 300 and 400 °C in vacuum for 0.5 hour for amorphous and crystalline materials, respectively.

Instead of synthesizing TiFe from elemental powders, Emami *et al.*[47] crushed and then ball milled a commercial TiFe ingot. Then, they exposed the powder sample to air for one month and before PCI analysis, activated it in vacuum at 150 °C for 2 hours. Readily, the sample absorbed 1.5 wt.% $H_2$. In comparison, the same crushed ingot exposed to air and only annealed did not absorb hydrogen, clearly showing the activation effect induced by ball milling.

Instead of ball milling, Edalati *et al.*[46] used mechanically activated TiFe by high pressure torsion. Small TiFe disks were pressed under 60 kbar in air, and then annealed in vacuum at 150 °C, for 2 hours. The resulting sample stored 1.7 wt.% $H_2$ during the first hydrogenation. Later, the same group investigated the effect of groove rolling on TiFe previously activated by high pressure torsion.[49] This latter sample required a few absorption/desorption cycles before reaching a capacity of 1.7 wt.% $H_2$. However, after air exposure for one day, it remains activated showing the same hydrogen uptake characteristics in the subsequent cycling.

In conclusion, easy activation in TiFe intermetallic compound can be promoted by a mechanical treatment or by the formation of secondary phases. The latter can be attained varying the Ti/Fe ratio with the precipitation of β-Ti or $TiFe_2$ for Ti-rich and Fe-rich alloys, respectively.[86–89]

**Thermodynamics of hydrogen sorption**

The first Pressure Composition Isotherm (PCI) curves of the TiFe-H system were monitored by Reilly *et al.*[68]. An example of absorption/desorption PCI isotherm at 40 °C is displayed in **Figure 4**. Three different regions were observed during the absorption of hydrogen: a steep pressure increase at low H-content (< $TiFeH_{0.1}$), followed by two pressure plateaus located at $P_{H_2}$ = 1.5 and ~40 bar and extending from 0.1 < H f.u.$^{-1}$ < 1, and 1 < H f.u.$^{-1}$ < 2, respectively. The initial branch (0.1< H f.u.$^{-1}$) was associated with the formation of the α-solid solution. The first plateau was attributed to the phase transition from the α-phase into the β-monohydrides. The second plateau, which is rather sloppy, was ascribed to the γ-dihydride formation.

In **Figure 4**, the length of the first plateau differs between absorption and desorption, suggesting different hydrogen contents for the intermediate β-phases. As mentioned above, Schefer *et al.*[62] proposed the existence of two different $\beta_1$ and $\beta_2$ phases, with similar crystal structures, except for small differences in the occupancy of the octahedral sites. The occurrence of these phases has been further investigated with volumetric measurements by Reidinger *et al.*[90]. On absorption, only $\beta_2$-$TiFeH_{1.4}$ was observed, while during desorption both $\beta_1$-$TiFeH_{1.0}$ and $\beta_2$-$TiFeH_{1.4}$ were detected. Based on these results, they suggested that the formation of the β-phases is related to the presence of strains induced by the absorption of hydrogen. This assumption was later confirmed by Reilly *et al.*[91], who, after activation, obtained a free-strain sample by annealing overnight at 800 °C under helium, and they achieved the full hydrogenation state in a single $\alpha \rightarrow \gamma$ step, without detecting any β-phase during absorption. However, after the formation of the γ-phase, which induces a volume expansion and thus lattice strains, β-phases appeared again during desorption. In addition, they demonstrated the strain effect on the overall performance of TiFe while cycling. They observed a decrease of the quantity of absorbed hydrogen with the increase of cycle number, mainly due to the disappearance of the upper γ-phase plateau, which shifts to higher pressures. On the other hand, the lower β-phase plateau seems unaffected. After several cycles, the quantity of hydrogen reaches a steady state value, suggesting a saturation of the internal strain. By annealing the samples for 2 days at various temperatures (from 230 to 350 °C), thus reducing the strain, some capacity was recovered, observing again the formation of the γ-phase.

Hydrogenation of amorphous TiFe showed no plateau pressure and low quantity of absorbed $H_2$ (0.3 wt.%) while nanocrystalline (5 nm size) TiFe displayed a single plateau with higher hydrogen content (0.9 wt.%).[32] Haraki *et al.* prepared TiFe from the elements by two different techniques: mechanical alloying and radio frequency melting.[38] After synthesis, the melted sample was later ball milled for 5 hours, and both TiFe specimens were annealed in vacuum for 2 hours at 300 °C before hydrogen absorption analysis. Interestingly, both samples exhibited absorption/desorption plateaus at lower pressures compared to TiFe produced by conventional arc melting. However, the PCI curves differ in shape and quantity of absorbed $H_2$. TiFe prepared by ball milling absorbed 1.3 wt.% exhibiting a single plateau, whereas the one prepared by radio frequency melting clearly showed two plateaus, reaching a content of 1.7 wt.% $H_2$, and suggesting the formation of both β and γ phases. The disappearance of the γ-phase formation in ball milled TiFe was confirmed by Zadorozhnyy *et* *al.*[92]. After an activation at 400 °C under 10 bar $H_2$ for 0.5 hour, a single plateau was found for absorption. X-ray diffraction analysis after hydrogenation (at 1.1 wt.% $H_2$) showed that only the monohydride β was formed.

By monitoring PCI curves at different temperatures, thermodynamic parameters can be determined thanks to the Van't Hoff equation:

$$\ln\left( {}^{P_p}/_{P^\circ} \right) = \frac{\Delta H}{\mathrm{R}T} - \frac{\Delta S}{\mathrm{R}}$$

where: $P_\mathrm{p}$ is the equilibrium plateau pressure (atm), $P^\circ$ the standard pressure (1 atm), R the gas constant (8.314 J $mol^{-1}$ $K^{-1}$), $T$ the temperature (K), $\Delta H$ the enthalpy change (J $mol.H_2^{-1}$), and $\Delta S$ the entropy change (J $mol.H_2^{-1}$ $K^{-1}$). It is worth to note that, due to hysteresis effects, enthalpy and entropy values evaluated by the Van't Hoff plot can differ on absorption and desorption.

The first thermodynamic data for hydrogen sorption in TiFe were reported by Reilly *et al.*,[68] providing values of $\Delta H_{1st}{}^{d}$ = 28.1 kJ $mol.H_2^{-1}$ and $\Delta S_{1st}{}^{d}$= 106 J $mol.H_2^{-1}$ $K^{-1}$, during $\beta \rightarrow \alpha$ and $\Delta H_{2nd}{}^{d}$ = 33.7 kJ $mol.H_2^{-1}$ and $\Delta S_{2nd}{}^{d}$= 132 J $mol.H_2^{-1}$ $K^{-1}$, during $\gamma \rightarrow \beta$ desorption reactions, respectively.

Later, a more detailed investigation of the thermodynamics of both hydrogen absorption and desorption reactions in TiFe was performed by Wenzl *et al.*[93]. Slight differences were found between absorption/desorption due to hysteresis loop. It is interesting to notice that, during the hydrogen absorption, the transformation α to β for the first plateau is less exothermic than that of β to γ for the second plateau (i.e. -25.4 kJ $mol.H_2^{-1}$ and -29.8 kJ $mol.H_2^{-1}$, respectively),[93] which is unusual in multi-plateau systems.[54,66,68,93]. In fact, if the entropy change is assumed to be constant (typically 130 J $mol.H_2^{-1}$ $K^{-1}$ as result of the entropy change of hydrogen from the gas phase into the solid state of the hydride). However, enthalpy evaluation from PCI data were confirmed by calorimetric analyses, which allow a direct measurement of the heat of reaction ($Q$), hence the enthalpy changes ($\Delta H = -\Delta Q$).[43,93] Results of thermodynamic analyses of hydrogen sorption reactions in TiFe, performed by both PCI measurements and calorimetric experiments, are summarized in **Table 1**. It is observed that the entropy change in TiFe is anomalously low (99 J mol.$H_2^{-1}$ $K^{-1}$) for the first plateau. Likely, this is linked to the high strains that stabilize the beta phase as mentioned above.

**Kinetics of hydrogen sorption**

The kinetics of hydrogen sorption in TiFe was first investigated by Park *et al.*,[94] to determine reaction rates, mechanisms and rate-limiting steps. As shown in **Figure 5**, hydrogen absorption rates were determined as a function of the reacted fraction, showing a maximum at ~ 25% of reaction that evidences two different mechanisms. They were initially ascribed to nucleation and growth, at the start of reaction, followed by hydrogen diffusion through an enveloping hydride layer after the rate maximum. However, the authors doubted about the first step assignment due to too fast absorption rates. Through a careful analysis of the absorption rate as a function of the gas pressure, they noticed that, before the maximum, it increases linearly, suggesting a step controlled either by $H_2$ mass transfer through cracks or surface chemisorption. SEM images of activated TiFe showed very large cracks facilitating hydrogen transport; therefore, chemisorption was suggested as initial rate-controlling step. Park *et al.* [94] proposed a core-shell model to explain observed kinetics, where the hydrogenation reaction proceeds as follows: hydrogen is chemisorbed on TiFe surface, from which the nucleation and growth of the hydride occurs, and then hydrogen slowly diffuses through the hydride layer in the last step of hydrogenation. Furthermore, Bowman *et al.* studied hydrogen sorption kinetics by NMR measurement to determine hydrogen diffusivites and activation energies at a local (microscopic) scale.[95–98] Compared to other metallic hydrides, which generally exhibit at room temperature a hydrogen diffusion coefficient in the range $10^{-6}$ – $10^{-8}$ $cm^2$ $s^{-1}$,[99] Bowman *et al.* found a value of the order of $10^{-12}$ $cm^2$ $s^{-1}$,[95], for β-TiFeH. This slow diffusion was attributed to the ordered structure restricting possible diffusion path, since H1 sites are almost fully occupied, while only a few H atoms are located in H2 sites.

**Cycling and resistance to poisoning**

One major fact to consider for practical application is the alloy degradation when cycled for long periods. Changes in the PCI curves of TiFe after cycling were reported by Goodell *et al.*[100]. Freshly activated TiFe exhibited two plateaus for the formation of β-monohydrides and γ-dihydride, with large hysteresis between absorption and desorption. However, with the increase of cycle numbers, the hysteresis gap decreases, but the γ-phase plateau shifted towards higher-pressure values, until it disappeared. Similar results were found also by Reilly *et al.*,[91] who showed that the PCI curves change in shape during several cycles, until they stabilize becoming almost independent on the cycle number. The authors supposed that, until lattice strain and defects do not reach saturation, the isotherms keep changing. This implies that the disappearance of the γ-phase is due to the presence of internal stress and defects, due to an expansion and shrinking of the unit cell during hydrogenation and dehydrogenation, respectively. Further analysis performed by Ahn *et al.*[101] confirmed the reduction of hydrogen stored due to the disappearance of γ-dihydride because of stress. Moreover, they observed a decrease also in the hydrogenation rate with the number of cycles. The authors suggested that, besides lattice distortion, also the formation of stable hydrides ($TiH_x$) due to alloy disproportionation during cycling could be a cause of the degradation. Indeed, stable hydrides do not release hydrogen, and their formation hinders hydrogenation on the TiFe surface due to rearrangements of neighbour atoms and the introduction of lattice strain.

Besides cycling-induced degradation, contaminants in the hydrogen gas such as $H_2O$, $O_2$, $CO_2$ and CO have a prominent influence. Adsorption of impurities at active sites on TiFe surface will prevent hydrogen molecules to dissociate during the chemisorption step. As demonstrated by Sandrock *et al.*[102], this passivation is generally manifested as a decrease in the reaction rate or a reduction in the storage capacity. These authors have investigated the effect of $H_2$ containing 300 ppm of $H_2O$, $O_2$ and CO on the cycling of TiFe. $H_2O$ and $O_2$ split on the surface, forming a thick passivating layer composed by complex oxides. The effect of this layer is similar in both cases, exhibiting a continuous decrease in the quantity of hydrogen stored during cycling. The main observed difference is that $O_2$ reacts faster than $H_2O$ at the surface. In both cases, TiFe could be partially reactivated cycling at moderate temperature (80 °C) with pure $H_2$. On the other hand, CO has shown to be more detrimental than $H_2O$ and $O_2$. It is adsorbed in less than one minute, completely deactivating TiFe in a few cycles. However, TiFe poisoned by CO was easily reactivated by simply cycling at room temperature under pure $H_2$. Additional information was provided by Block *et al.*[103], who investigated also the effect of $CO_2$, $CH_4$ and $H_2S$ in various concentrations. The presence of 10 vol.% $CH_4$ showed a stable and slight decrease in the capacity and reaction rate of TiFe. Surprisingly, when pure $H_2$ was provided again, the active material exhibited a reaction rate even faster than before, restoring also its hydrogen absorption capacity. The authors suggested that $CH_4$ does not passivate TiFe, and the decrease during cycling was probably due to the lower $H_2$ partial pressure in presence of methane. In the presence of $CO_2$ there is a constant decrease in the storage capacity during cycling and, moreover, the absorption rate decreases with the increase of impurity concentration into $H_2$. A concentration of 1 vol.% $CO_2$ in the gas stream was enough to fully passivate TiFe after two cycles. The sample was reactivated by cycling with pure hydrogen at 127 °C. Introducing 0.2 vol.% $H_2S$ did not affect the reaction rate, but it strongly reduced the quantity of hydrogen absorbed upon cycling, so that few cycles were enough to completely deactivate TiFe. Even performing intensive heat treatments, the authors were not able to reactivate the sample due to the presence of a stable sulphur layer on the surface of TiFe, which inhibited the absorption of hydrogen.

From these results, it can be concluded that TiFe hydrogenation properties easily deteriorate in the presence of contaminants. To face this issue, two main strategies were suggested: the design and implementation of reactivation systems or the enhancement of TiFe resistance to poisoning. Resistance to passivation reaction might be induced by adding a secondary phase, but still no complete resistance to contamination has been reported in the literature for TiFe. Hence, leaks and gas purity must be carefully checked for long-cycling applications for hydrogen storage.

All properties mentioned above for binary TiFe can be tailored by chemical substitutions and this topic will be discussed in detail in the following sections.

## Modifications of TiFe properties by substitutions

Extensive studies have been performed to synthetize and characterize substituted TiFe intermetallic compounds with many elements, as it can be visualized in the periodic table reported in **Figure 2**. The substitution of Fe or of Ti has been the subject of recent papers that evidence the role of Ti-substitution or Fe-substitution and their effect on hydrogen storage properties.[104] Optimization of operational pressure range, a theoretical understanding of alloy thermodynamics, the role of secondary phases' formation or TiFe single phase domain compositional stretching need to be better considered in a full picture of available studies. Substitution can significantly lower plateau pressure or make full hydrogenation more difficult, decreasing the usable capacity. On the other hand, for example, Mn can change equilibrium pressure introducing a smoothing effect, levering plateau pressure in a narrow pressure range and maximizing the reversible capacity.

Vivid literature studies on substitutional effects have been carried out aiming to tailor hydrogenation properties of TiFe, indicating that substitution for Fe is dominant. In the following, mono-substituted system will be considered first, then we will specifically focus on the manganese-substituted system and finally prominent examples of Ti(Fe,Mn) multi-substituted alloys will be presented. Throughout the description of literature results, when studied, quaternary alloy are reported as well, while a focus on substitutional effect and Ti or Fe substitution are commented in detail in the discussion section. Few examples of reported additives or catalysts (as nanoparticles or oxides) will be cited and discussed too when relevant.

Substitutional elements are classified according to their location in the periodic table. Investigated TiFe-*M* systems, their hydrogen storage properties and thermodynamics are summarized in **Table 2**. In the case of single elemental substitution, an empirical geometric model was proposed by Lundin *et al.*[105] and Achard *et al.*[106], reporting that by enlarging the unit-cell volume of TiFe, interstitial holes size increases and plateau pressures in PCI curves shift to lower values. This empirical law, to which many intermetallic systems obey, can differ from that observed for some substitutions, therefore, electronic band structure should be considered and implemented with *ab-initio* studies, as demonstrated by Jung *et al.*[104].

**Alkaline earths (Mg, Be)**

Magnesium (Mg, $r_{Mg}$= 0.16013 nm, radius values reported from [107], for comparison $r_{Ti}$= 0.14615 nm and $r_{Fe}$= 0.12412 nm) can be substituted up to 2 at.% by ball milling, while up to 6 at.% the precipitation of Fe as secondary phase is observed.[108] It induces an easier activation compared to pure TiFe, an enlargement of the cell parameter and a concomitant decrease of equilibrium pressure in the PCI, which presents a single plateau related to the formation of the monohydride.[108] So, in the case of Mg substitution, the formation of the $\gamma$ phase is suppressed thus reducing the reversible capacity of the material.[108]

The substitution of Fe with Beryllium (Be, $r_{Be}$= 0.1128 nm), a smaller element with respect to Fe ($r_{Fe}$= 0.12412 nm), up to 15 at.%, evidences that geometrical factors alone fail to explain the variation of hydride stability. Although the TiFe unit-cell shrinks with Be substitution, the plateau pressures decrease as reported by Bruzzone *et al.*[109]. Furthermore, lower capacity but narrower hysteresis and sufficiently good kinetic were evidenced.[109] Besides, the thermodynamics are modified, evidencing higher values of $\Delta H$ introducing Be.[110]

**Early transition metals (Zr, Hf, V, Nb, Ta)**

Zirconium (Zr, $r_{Zr}$= 0.16025 nm) substitution for Ti ($r_{Ti}$= 0.14615 nm) has a positive effect on activation.[111,112] Following the geometric model, it increases the cell parameter of TiFe and decreases the plateau pressures.[113–117] However, a decrease of reversible capacity was observed and related to the enlarged solubility of hydrogen in the solid solution ($\alpha$ phase) at high Zr content.[113] Zr substitution leads to slopping plateaus, no variation in hysteresis and fast kinetics.[118] Jain *et al.* studied the effect of 4 wt.% Zr addition to TiFe, which confirms the positive effect of this substitution for activation (no need of thermal treatment), fast kinetics, a good maximum capacity of 1.60 wt.% at 20 bar and 40 °C, and a good resistance to air.[119] However, an increase in hysteresis was observed as well, in contrast with previous findings.[119] Mechanical treatment (i.e.

ball milling and cold rolling) can easily recover hydrogen capacity of this material after air exposure.[120]

Hafnium (Hf, $r_{Hf}$= 0.15775 nm) can be introduced into TiFe up to 2 at.%, causing an increase of the cell parameter and a subsequent decrease of plateau pressures. The formation of secondary phases have also been observed, improving activation (possible at room temperature and 20 bar) and kinetics, but slightly reducing the hydrogen capacity of the material.[121]

Vanadium (V, $r_V$= 0.1316 nm) can substitute both Ti and Fe in TiFe.[104,122] Furthermore, it has been reported that the addition of V to $TiFe_{0.90}$ decreases the total capacity of the material, but on the other hand it decreases hysteresis between absorption and desorption, even if slopped plateaus are observed.[123]. The addition of V levers the difference between the two plateaus introducing a smoothing effect towards a single plateau that has been widely discussed by Jung *et al.* combining DFT calculations and experiments.[104] They evidenced a stronger effect in lowering both plateau pressure when V substitutes Fe. While V substitution for Ti increases the first plateau pressure and decreases the second one.[104] However, V substitution does not improve kinetics, neither cycling stability or resistance to poisoning and oxidation.[124,125] Furthermore, it enlarges the cell parameter of the TiFe phase and promotes the formation of smaller crystallite sizes.[126]

Niobium (Nb, $r_{Nb}$= 0.1429 nm) substitution was studied by in-situ X-ray diffraction in $TiFe_{0.90}Nb_{0.10}$. It evidenced the formation of β-Ti as secondary phase that starts absorbing hydrogen upon first hydrogenation, allowing easy activation at 22 °C and under 50 bar after 5000 s of incubation time.[87] The incubation time can be shorten by the combined substitution of Ti by Nb and the addition of $Fe_2O_3$. This mixture results in the formation of secondary phases that improve the activation process (making the material more brittle). In addition, a slight shift of PCI curves towards lower plateau pressure values was observed.[127] The lower plateau pressure can be linked to the increasing lattice constant of TiFe introducing Nb.[128] Similar results have been observed recently in Nb-substituted materials by mechanical alloying, but evidencing the suppression of the γ phase, thus decreasing the total capacity.[129] The suppression of the second plateau by ball milling has been justified by nanostructuration, defects and possibly oxygen contamination, influencing and deforming the coordination site for hydrogen.[129] Improved resistance to poisoning was mentioned as well.[129]

Finally, Tantalum (Ta, $r_{Ta}$= 0.1430 nm) substitution was recently studied by Kuziora *et al.*[130]. It enlarges the cell parameter of TiFe and lowers the equilibrium pressure in PCI curves.[130]

**Late transition metals (Cr, Mo, Co, Ni, Pd, Cu)**

Chromium (Cr, $r_{Cr}$= 0.12491 nm) substitution in $TiFe_{0.90}Cr_{0.10}$ and $TiFe_{0.95}Cr_{0.05}$ forms $TiCr_2$ as a secondary phase, which helps in accelerating activation process of the alloys.[122,131] The Cr substitution stabilized as well the first plateau, while reducing the length of the second one. Cr-substituted TiFe alloys have higher hardness, are more brittle and easier to pulverise with respect to the non-substituted compound. This can be the reason for improved kinetics and reduced hysteresis due to Cr-substitutions.[132] On the other hand, this conclusion is in contrast with the higher strain claimed due to Cr substitution. As a matter of fact, usually hysteresis is generated either by elastic strain or by plastic deformation (dislocations, slip bands), and it increases with hardening.[133] By mechanochemistry, up to 6 at.% of Cr can be included into TiFe, enlarging the cell parameter with a small expansion of the cell volume, lowering crystallite size, and simplifying the activation process.[134] Differently to what is expected from geometric considerations, both plateau pressures are shifted to higher value compared to TiFe, and the gamma phase is also destabilised.[134]

The combined substitution of Chromium and Yttrium in TiFe evidenced an enlargement of the cell constant, the formation of secondary phases (Ti, Cr-Fe solid solution, α-Y), improved kinetic and sloped PCI curves, with a lowering of the plateau pressures and hysteresis.[132] On the other hand, the combined substitution of Cr and Zr in TiFe evidenced the formation of $TiFe_2$ as secondary phase, which acts as gateway for hydrogen, easing the activation process of the material.[135] The material was activated at 28 °C under 31 bar, and it did not lose any capacity after 50 cycles.[135]

Molybdenum (Mo, $r_{Mo}$= 0.13626 nm) substitution was reported to lower plateau pressures and to introduce sloppy plateaus.[130,136]

The substitution of Fe with Cobalt (Co, $r_{Co}$= 0.1251 nm) linearly decreases the first plateau pressure and also reduces the capacity of the material, since only the monohydride is formed.[137–139] Recently, improved resistance to poisoning and the suppression of the γ phase was also observed for Co-substituted materials by mechanical alloying, while, in the as-cast conditions, the second plateau was observed to increase the equilibrium pressure.[129]

Nickel (Ni, $r_{Ni}$= 0.12459 nm) substitutes Fe with no significant changes in the microstructure[122], improves activation and lowers hysteresis between absorption and desorption curves.[140] However, it increases the pressure gap between the first and second plateau, reducing the reversible capacity of the material in a narrow pressure range.[141] Owing to Ni substitution, improved kinetics is observed because of promoted surface sorption. In fact, the rate determining step in hydrogen sorption is the bulk reaction.[142] Furthermore, the use of catalysts like Ni nanoparticles at TiFe surface has been reported to considerably enhance the rate of hydrogenation process, even if it cannot be consider as a substituent.[138,143] Additionally, increasing the Ni content, the cell parameter of TiFe increases, decreasing plateaus pressures, decreasing capacity and increasing the decomposition temperature and the cohesive energy of the hydride.[144,145] Modified thermodynamics have been reported as well, with lower value of enthalpy and entropy for Ni-substituted TiFe.[146,147] Distorted γ region has been observed.[138]

Addition of Ni stabilized the monohydride as observed in mono-substituted $TiFe_{1-x}Ni_x$ compounds.[138,145] It improves cyclability of the material up to 65000 cycles and the reduced loss in capacity was related to possible hydrogen trapped or deactivated reaction site in the material ($TiFe_{0.80}Ni_{0.20}$).[148] Nevertheless, Jain *et al.* reported a general negative impact on hydrogen storage properties of Ni substitution for Fe.[119]

In the literature, few examples of multi-substituted TiFe-Ni alloys are reported. In the same paper cited before, Jain *et al.* reported a beneficial improvement of activation and kinetics by introducing 4 wt.% of $Zr_7Ni_{10}$. It reduces the plateaus pressure but decreases as well the capacity, down to 1.34 wt.% at 40 °C and 20 bar, thus with negative effect on hydrogenation properties.[119,149]

The simultaneous addition of Ni and V, or Ni and Nb, to TiFe was beneficial for the activation process, possible at 28 °C and under 20 bar with a short incubation time (30-40 minutes), reducing hysteresis, lowering the plateau pressures, and granting good capacity and kinetics with no sensible variation of the thermodynamics.[150] In contrast, a significant variation of the thermodynamics has been evidenced in the case of combined Ni and Mg substitution.[146]

Substitution in TiFe by alloying Palladium (Pd, $r_{Pd}$= 0.13754 nm) mitigates the activation process, lowers the plateau pressure of the monohydride (enlarging the cell parameter of TiFe) but it has no effect on the stability of the γ phase.[151] Beside, the addition of free Pd as nano-catalyst nanostructured with TiFe by milling has been reported in many studies to improve air resistance to poisoning and facile activation at room temperature.[152–156] Equally to Ni, the use of Pd nanoparticles as catalyst considerably enhances the rate of hydrogenation process.[138,143]Mechanochemical synthesis can introduce Copper (Cu, $r_{Cu}$= 0.1278 nm) into TiFe, enlarging the cell parameter, thus reducing the first plateau pressure of the binary compound.[117] The combination of Cu substitution for Fe in TiFe and the addition of $Fe_2O_3$ has a positive impact on activation process, that is promoted thanks to a more brittle TiFe matrix, an enhanced formation of active surface by cracking, and the lowering of the plateau pressure.[127]

**Rare-earths (Y, La, Ce, Mm)**

Yittrium substitution (Y, $r_Y$= 0.18015 nm) into TiFe modifies the properties of the material. The increase in Y content linearly increases the cell parameter of TiFe and reduces the crystallite size, without changing significantly the microstructure.[157] A fast kinetics and an easy activation process are observed, with no incubation time at room temperature and 25 bar.[157] However, on increasing Y content, the capacity of the material decreases, due to the formation of secondary phases ($TiFe_2$, Y and Ti precipitates).[157]

The addition of 5 wt.% of Lanthanum (La, $r_{La}$= 0.1879 nm) to TiFe faces the issue of La immiscibility in the intermetallic compound, but still improves the activation process, thanks to crack formation.

Indeed, the incubation time is reduced increasing La content.[158] The capacity is claimed to be improved and small hysteresis is observed.[158]

Cerium (Ce, $r_{Ce}$= 0.18247 nm) substitution for Fe improves activation process and kinetics, because of the formation of small crystallite sizes, that induce high surface reactivity, while increasing the cell parameter of TiFe.[159]

The addition of Mischmetal (*Mm*, containing La, Ce, Pr, Nd) to TiFe allows easy activation at room temperature after a short incubation time, owing to the cracking of the material caused by *Mm* inclusions. No evidence of *Mm* substitution in TiFe were clearly reported.[160,161] In these materials, TEM analysis evidenced the formation of channels that could be depicted by electron micrograph in the hydrogenated sample, and that improves absorption kinetic.[162] Xin-Nan *et al.* demonstrated that the addition of Mischmetal to TiFe enhances the resistance towards impurities (mainly $O_2$ and $CO_2$).[163] In their work, TiFe with 3 wt.% *Mm* was cycled under hydrogen with a purity below 99%. Moreover, they claimed that the material exhibits a lower decrease in capacity during cycling compared to pure TiFe. When exposed to pure $H_2$, it recovers its capacity within a few cycles without any annealing treatment.

**p-block elements (Al, Si, Sn)**

The effect of Aluminium (Al, $r_{Al}$= 0.14317 nm) substitution in TiFe has been extensively studied, evidencing an improvement of the kinetics, but the generation of sloped PCI curves.[164] Aluminium substitutes Fe increasing TiFe cell parameter.[122] It causes negative impact on hydrogenation properties. An increase of the Al content increases the slope of the PCIs, which, compared to TiFe, are shifted to higher pressure values, even if larger cell parameters are observed. [138] The second plateau either disappears or moves to high pressures, inhibiting the formation of the γ hydride, with a consequent drastic fall of the hydrogen capacity of the material.[110,164] The generation of a sloped plateau has been related compositional inhomogeneity in as cast samples and the formation of octahedral interstice's size gradient.[138] Increasing Al substitution for Fe also modifies the thermodynamics and reduces hysteresis, due to the difference in valence electrons between Al and Fe.[164] In the same study of Zadorozhnyy *et al.* cited before, up to 20 at.% of Al were substituted to Fe in TiFe by mechanochemistry, with results similar to those observed for Cr substitution.[134] Silicon (Si, $r_{Si}$= 0.1153 nm) can substitute Fe, causing the formation of secondary phases such as $TiFe_2$ and Ti, depending on the stoichiometry, and diminishing the hydrogen storage capacity.[122] Si has deleterious influence on the capacity because the second plateau is shifted to high pressures, while the first plateau becomes very sloppy and shifted to lower pressure as compared to TiFe, as it occurs for Cu and Ni substitutions.[122]

Kulshreshtha *et al.* studied Tin (Sn, $r_{Sn}$= 0.162 nm) that substitutes both Ti and Fe in TiFe. Activation and kinetics are improved, while capacity is decreased.[165] The improvement of activation has been related to the formation of $TiFe_2$ as secondary precipitates (in a critical minimum size), which causes strain induced micro cracks owing to different thermal expansion. Surprisingly, the Sn substitution causes a shrink in cell volume of TiFe, leading to an increase of pressure for both plateaus, together with an increase of corresponding $\Delta H$ and $\Delta S$ values.[165]

**Reactive non-metal elements (B, C, N, O, S)**

In general, reactive non-metal elements are present as interstitial atoms or promote the formation of secondary phases.

Small quantities of Boron (B, $r_B$= 0.082 nm) and Carbon (C, $r_C$= 0.0773 nm) induced the formation of secondary phases (i.e. $TiFe_2$ and Ti), promoting easy activation, but reducing drastically the storage capacity.[166] Sloped plateaus at high equilibrium pressures were observed, with no formation of the $\gamma$ phase.[166] Furthermore, Carbon and Nitrogen (N, $r_N$= 0.075 nm) form carbides and nitrides lowering the total capacity of the material.[122]

In 1977, Sandrock *et al.* discussed the effect of element contamination by Al, Si, C, N, O during material production and processing to phase homogeneity and microstructure of TiFe.[122] Oxygen (O, $r_O$= 0.073 nm) mostly forms oxides that causes capacity deterioration but could also help in activation.[122] Extensive studies have been dedicated to the understanding of oxygen influence in TiFe materials and the role of oxide phases in activation process, which however are not the focus of this review. Some related studies can be found in the references cited hereafter.[79,80,83,84,102,122,153,154,167–175]

The addition of Sulphur (S, $r_S$= 0.102 nm) to TiFe affects cycling properties, avoiding pulverisation of the material, improving activation at moderate temperature and reducing the incubation time, thanks to the formation of $Ti_2S$ as a secondary phase at the grain boundaries.[176] With the increase of S-content, a slight rise of plateau pressure in PCI is observed.[176] However, the addition of 1 at.% of S by ball milling evidenced an enlargement of the TiFe cell parameter, resulting in a decreased plateau pressure, and reduced reversible capacity, because of the suppression of the dihydride.[108] Furthermore, it has been reported that the introduction of small amount of sulphur in pure TiFe can improve the resistance to poisoning.[176]

**Manganese-substituted TiFe alloys**

Mn-substitution is of paramount importance in the design of TiFe alloys for large-scale storage application due to the improvement of the main hydrogenation properties. In addition, under the European strategy, Manganese (Mn, $r_{Mn}$= 0.135 nm), such as Fe, is inexpensive and is not listed as CRMs. Many studies can be found in the literature regarding Mn substitutions in TiFe, together with determined thermodynamic properties as reported in **Table 3**.

The ternary Ti-Fe-Mn phase diagram presents many phases at 1000 °C, as reported in **Figure 6**. The β-Ti solid solution region, as well as the $Ti(Fe,Mn)_2$ C14 laves phases, have large homogeneity domains. At 1000 °C, the intermetallic compound TiFe exists in the range of 49.7 to 52.5 at.% Ti. Mn can substitute Fe in this compound up to 27 atomic percent in a narrow region of Ti composition, as reported by Dew-Hughes *et al.*[177]. For this reasons, the authors recently investigated Mn-substituted alloys in a wide range of composition, variating both Mn and Ti content.[178]

Manganese substitution for Fe enables the hydrogen sorption at lower pressure by enlarging the cell volume of TiFe.[70] The higher the Mn content, the lower is the hydrogen sorption pressure for both plateaus. Furthermore, easy activation of Mn-substituted TiFe has been related to highly reactive grain boundaries induced by segregation of metal atoms or cluster-like precipitates formation, that can deviate the concentration ratio of components especially at the surface.[70,84,179,180]

Reilly *et al.* were the first group in the 70's studying TiFe for hydrogen storage. They have published several reports on Mn substitution either in equiatomic or Ti-rich TiFe alloys.[79,88,105,136,181,182] They evidenced many positive improvements compared to pure TiFe. Mn modifies the microstructure, reduces the hysteresis,[140,183] improves the activation.[122] In addition, Mn substitutions promote the presence of secondary phases, such as β-Ti solid solution or $Ti(Fe,Mn)_2$, that facilitate the alloy cracking and the creation of fresh clean surfaces during the first hydrogen absorption. Furthermore, good long-term cycling performances (without disproportionation or phase separation) and improved hydrogen capacity are reported.[105,136,181] The latter probably relates to the thermodynamic stabilisation (i.e. decreased plateau pressure) of the dihydride, so that hydrogen saturation in the γ-phase can be easily reached at low applied pressures.[105,136,181] It was also evidenced that a high amount of Mn (e.g. $TiFe_{0.70}Mn_{0.30}$) actually decreases the total capacity of the material, due to the significant formation of secondary phases, *i.e.* $Ti(Fe,Mn)_2$, which are not reactive to hydrogen in mild condition of temperature and pressure.[136]

Lee *et al.* reported better kinetics and activation when Mn is introduced into TiFe, thanks to the presence of secondary phases, improving capacity as well.[131] In their study, the PCI of $TiFe_{0.90}Mn_{0.10}$ and $TiFe_{0.80}Mn_{0.20}$ reported at 50 °C showed lower plateau pressures and reduced dihydride region due to the substitution, compared to TiFe.[131] Since, by increasing the Mn content, an improvement of kinetics is observed, activation is realized in short incubation time under moderate conditions (room temperature and low pressure). Furthermore the hysteresis between absorption and desorption is reduced.[52] The good kinetics seems not to be related to hydride stability nor to particle size, but to the formation of Mn clusters that enhances a faster hydrogenation compared to TiFe.[184]

Sandrock *et al.* reported the thermodynamics of $TiFe_{0.85}Mn_{0.15}$, where the incorporation of Mn in the compound generates higher enthalpy and entropy of reaction, but also higher capacity, compared to TiFe.[185] $TiFe_{0.85}Mn_{0.15}$ was also recently investigated, evidencing an easy single-step reactivation at 300 °C after oxidation in air.[186]

Milling effect in Ti(Fe,Mn) alloys evidenced that a reduction of size and microstructure promotes an easy activation and an enhanced kinetics, slightly modifying hydrogenation properties and generating a sloping plateau, which is stabilized at lower pressure compared to the pristine alloy.[117,187]

Severe plastic deformation, such as high pressure torsion, has been used to improve activation and air resistivity on Mn-substituted TiFe alloys, owing to the formation of lattice defects at the grain boundaries and amorphous regions, that are claimed to act as channels for fast hydrogen diffusion facilitating activation.[48]

Lee *et al.* stated that the addition of manganese to TiFe increases the hydriding rate.[188] In $TiFe_{0.80}Mn_{0.20}$, at low reacted fraction, the rate-determining step is chemisorption, while, towards the end of sorption, it is the chemical reaction at the metal-hydride interface.[188] The hydriding reaction rate increases with increasing pressure at constant temperature and with decreasing temperature at constant pressure. In fact, if the temperature increases at constant pressure, the exponential term of the rate equation, which includes the activation energy term, increases, but the equilibrium pressure or the driving force term decreases. Thus, since a relatively small activation energy is compared to a rather drastic pressure change with temperature, it results in a decrease in the reaction rate with temperature.[188] Lee *et al.* calculated the rate constant for TiFe and $TiFe_{0.80}Mn_{0.20}$ and the obtained values suggest chemisorption as main rate-determining step. The rate constant increases through Mn for Fe substitution.[188]

Another study confirmed that the rate determining step is the reaction of hydrogenation at the surface, which is followed by that in the bulk.[142] The latter becomes dominant at lower temperatures and at the later stages of reaction.[142] The study considered also Ni-substituted TiFe alloy ($TiFe_{0.90}Ni_{0.10}$), in which the surface reaction is no longer found to be the dominant kinetic mechanism.[142]

Reilly *et al.*[105,189] and Challet *et al.*[141] reported some studies on Mn-substituted TiFe for Ti-rich alloys (i.e. $TiFe_{0.70}Mn_{0.20}$). Reilly *et al.* monitored a PCI curve at 30 °C with strongly slopped and short plateaus and with reduced hysteresis as well [105,189]. The shape of PCI curve differs from that reported by Challet *et al.*,[141] which presents the typical double flat plateaus. The materials from Challet *et al.* were annealed at 1000 °C for one week, thus, possible differences in PCI curves could be related to different homogeneity in chemical composition.

Guéguen *et al.*[123] investigated as well the Ti-rich $TiFe_{0.80}Mn_{0.10}$ composition. In their samples, Ti-type and $Ti_2Fe$-type precipitates were observed as secondary phases favouring alloy activation without any thermal treatment.

Recently Mn substituted TiFe materials have been scaled-up and produced by some industrial companies. Bellosta Von Colbe *et al.* demonstrated that 6 kg of an industrial Ti(Fe,Mn) alloy, containing 6 at.% Mn, could be easily activated at a large scale after a short ball-milling treatment. This treatment reduced the particle and crystallite size, improving activation without thermal treatment (at RT and $P_{H_2}$ = 20 bar), kinetics, cyclability and hydrogenation properties.[190]

However, even if activation is improved and is reproducible in Mn-substituted TiFe compounds, they still suffer from sensitivity to contaminants such as $O_2$, CO, $CO_2$, $H_2S$, $H_2O$[136,168] and $Cl_2$.[183] Shwartz *et al.* reported that Ti is oxidised by $O_2$ to form $TiO_2$ in two steps. $H_2O$ directly reacts with Ti(Fe,Mn) alloy to form $TiO_2$, whereas Fe is not oxidised and Mn is oxidised only by oxygen and not by water.[175] Mn acts as preferable oxidation element.[175,179]

Another example was provided in the work of Sandrock *et al.*,[102] where iron was partially substituted by manganese, forming the $TiFe_{0.85}Mn_{0.15}$ alloy by arc melting. This compound has a similar behaviour to pure TiFe in presence of $H_2O$ and $O_2$, but a higher resistance towards CO, with a lower decrease in the storage capacity. Moreover, its reactivation after being exposed to CO was completed within the first cycle under pure $H_2$ at room temperature, whereas TiFe needed several absorption/desorption cycles.[102]

**Multi-substituted Ti(Fe,Mn) alloys**

Thanks to the good properties of Mn-substituted Ti(Fe,Mn) pseudo-binary intermetallics, further research has been focused on their tailoring adding further elements and studying the synergic effect with Mn on hydrogenation thermodynamics. Main results are summarized in **Table 4**.

Ball milling has been used to incorporate both Zr and Mn into TiFe intermetallic compound, enhancing the activation of the material, and making it possible at room temperature and 40 bar.[191] By increasing the content of Mn and Zr, the equilibrium pressure is lowered, while activation, kinetics and resistance to air are improved.[192,193] In contrast, capacity is reduced, because only the monohydride can be formed by Zr substitution.[194]

The introduction of both V and Mn results in lower plateaus pressures, decreased hysteresis, flatter plateaus, good activation and capacity, showing a synergic effect of both Mn and V.[123] The effect of the double substitution in Fe-rich alloys (detailed composition reported in **Table 4**) was explored by Mitrokhin *et al.* evidencing the formation of a secondary *C*14 Laves phase.[195,196] Combined Mn and V substitution reduces the pressure gap (V effect) between the two plateaus, that merge into one at lower pressure compared to pure TiFe, while enthalpy of reaction increases (Mn effect).[195] Even if resistance to contaminants was not improved, easy activation was observed but with very slow kinetics.[195,196] $TiFe_{0.80}Mn_{0.20}$ with the addition of V was scaled up to 55 kg by Japan Metals & Chemicals Co. Lts and studied by Endo *et al.*, showing easy activation under 10 bar and 80°C.[197] Substitution effect of Co over Mn has been explored by Qu *et al.*[198]. An increase of the Co content decreases the cell volume of TiFe, while improving activation, resistance to pulverisation upon cycling and increasing capacity. As a drawback, PCI curves are sloped and the second plateau is observed at an increased equilibrium pressure.[198]

Another example of bi-substituted compounds is the Ni addition to pseudobinary Ti(Fe,Mn) compounds. Ni for Mn substitution has been evaluated, evidencing no significant variation of the cell parameter. Challet *et al.*[141] introduced both Mn and Ni into Ti-rich $TiFe_{0.90}$ observing that Ni addition improved activation process but strongly affects the hydrogenation properties as well. A decrease of the first plateau pressure and an increase of the second one was observed, stabilising the monohydride and destabilising the dihydride. As a consequence, the adjustment of both plateau pressures in a narrow pressure range is much more difficult to achieve when substituting Mn by Ni. Finally, the reversible capacity decreases with Ni content because of the increase of the pressure gap between the two plateaus.

Furthermore, the process history of the sample influences the shape of the PCI curves. For example, milling under argon with small amount of Ni, dispersed as a catalyst, allows to synthetize a material that does not need activation and has a longer cycle lifetime compared to pure TiFe.[199]

The combined substitution of Cu and Mn for Fe in $TiFe_{0.90}$ was recently evaluated by Dematteis *et al.*[200]. Easy activation and fast kinetics were granted thanks to the Mn substitution and the formation of small amount of secondary phases (β-Ti and $Ti_4Fe_2O$-type as precipitates). Cu augments the secondary phase amounts while increasing the TiFe cell parameter and decreasing the first plateau pressure. Similarly to Ni, a negative effect of Cu substitution is that it rises the second plateau pressure, revealing the predominance of electronic effects associated with this substitution that should be verified and deepen by *ab-initio* calculations of their electronic structure (i.e. analysis of density of states).

The combined substitution of Mn and Y causes an enlargement of the cell constant, thus lowering the plateau pressure in PCI curves, which are still flat and shows double plateaus only in desorption.[132] Y is generally not highly soluble into TiFe, so α-Y precipitates are formed.[132]

Addition of Ce to $TiFe_{0.90}Mn_{0.10}$ evidenced no effect on the thermodynamics and cycling properties of the material, while, increasing Ce content improves kinetics but slightly lower the hydrogen capacity.[201] The activation process is remarkably improved as well, requiring no annealing at high temperature and no incubation time at 80 °C and 40 bar.[201]

The study of $Ti_{1+x}Fe_{1-y}Mn_yMm_z$ ($x$ = 0.0-0.9, $y$ = 0.04-0.2, $z$ = 0.002-0.028) system evidenced that 0.5-1.0 wt.% of Mischmetal is the optimum amount for improving activation at room temperature in this system, preventing oxidation during processing, and slightly affecting the hydrogenation properties. In fact, at 49 °C the PCI curve presents a single slightly sloppy plateau.[202]

Recently, even three-substituted alloys have been explored as in the case of $TiFe_{0.86}Mn_{0.1}Y_{0.10-x}Cu_x$ ($0.01 \leq x \leq 0.09$).[203,204] Ali *et al.* showed that by increasing Y content, while decreasing Cu, the cell parameter of the cubic phase and the capacity are increased, while the plateaus pressure decreased. The formation of secondary phases (α-Y, CuY, $Cu_4Y$, $Cu_2Y$) likely allows easy activation. In their study, a complete determination of the PCI at different temperatures allowed the determination of the thermodynamics of these alloys.[203,204]

## Discussion and correlations

The definition of A or B-type atoms in terms of formation enthalpy of binary hydrides[60,205,206] is shown in **Figure 1**. In addition, the definitions of A- and B-type elements from the structural point of view can include geometric and electronic parameters. For instance, sigma phases in binary systems typically have an A and B element. The A element is poor in *d*-electrons, it has a bcc crystal structure, a large atomic radius and a preference for sites with large coordination numbers. Whereas the B element is rich in *d*-electrons, it has fcc or hcp crystal structure, a small atomic radius and a preference for sites with limited coordination number.[207]

Based on these definitions, elemental substitution for either Ti (A-type) or Fe (B-type) in TiFe intermetallic compound can be explained and the definition based on binary hydride enthalpies of formation is particularly effective (**Figure 1**). Unfortunately, in the performed literature survey, the location of the substitution element either at Ti or Fe sites is not always described and verified experimentally. **Table 5** shows an overview of elemental substitutional effects and it reports Ti or Fe substitution when known.

As a matter of fact, alkaline earths (Mg, Be) are expected to substitute Ti in the case of Mg (not verified experimentally), and Fe in the case of Be. Early transition metals possibly substitute Ti as they are A-type elements. However, Hafnium and Tantalum substitutions were not clearly investigated. Late transition metals substitute Fe as they are B-type elements. This behaviour has been verified for all elements except for Pd, which either has been added as a catalyst or substituted without further details on atomic site location. Rare earths possibly substitute Ti as they are A-type compounds. Though this is the case for Yttrium, Lanthanum and Mischmetal are reported to have no solubility in TiFe, and Cerium probably substitutes Fe. *p*-block elements are expected to substitute Fe, as they are B-type elements. However, in the case of Tin and Vanadium, they have been reported to have both A and B-type behaviour, as regards to substitution for Ti and Fe.

The extensive literature survey performed in this review evidences that, generally, the addition of a third element to TiFe intermetallic compound has beneficial effects on its activation, hydrogenation kinetics, cycling and resistance to poisoning properties. However, also some negative effects have been reported.

Addition of elements in the formulation of TiFe can lead either to a substitution (*i.e.* entering into the TiFe structure at Ti or Fe sites) or to an addition (*i.e.* forming precipitates like second intermetallic phases or oxides). An element can present both behaviours, depending on its limit of solubility in the TiFe phase. For the first case (substitution), the lattice parameter of TiFe phase is modified according to the change in its composition. It leads to modifications of the thermodynamic properties, such as enthalpy, entropy, PCI shape and number of plateaus. For the second one (addition), secondary phases are formed, often accompanied by microstructural changes and increased brittleness. It leads to modification on dynamic and ageing properties like activation, kinetics, poisoning, and cycling.

In this section, the review focuses on the effects of an elemental substitution or addition on properties of the TiFe intermetallic compound, which are essential to be known for efficient tailoring of this material towards real applications. The chemistry behind TiFe-substituted compound will be described together with some useful correlations and guidelines, highlighted by the extensive study of the literature.

**Activation and kinetics**

The introduction of a substitutional element in TiFe alloys always brings the positive effect of improving activation process and kinetics. It is reported that the TiFe activation is facilitated by adding Al, Si, Mn and Mg, which prevent oxidation, hence allowing the hydrogen absorption process. Cu and Ni substitution seems to be less effective compared to the previous cited elements.[169] The prevented TiFe oxidation by Al, Si, Mn and Mg may be related to the fact that these elements form more stable oxides than Cu and Ni, getting preferentially oxidized compared to TiFe, and behaving as oxygen getters.

Improved activation can be related to a catalytic behaviour of the additive or the formation of secondary phases at the grain boundaries, with enough size and abundance. Surface solid-gas interactions are enhanced by the presence of secondary phases or elemental clusters, which are highly reactive towards hydrogen, thus promoting hydrogen chemisorption and causing cracks by expansion during hydriding reaction. This process creates preferential channels for hydrogen towards the TiFe-phase or fresh oxide-free surfaces accessible for hydrogenation. Even if secondary phases have different mechanical behaviours, the activation process can be improved owing to the cracking process caused by different thermal expansion compared to TiFe. Alternatively, activation can also be facilitated by the creation of lattice defects (such as vacancies, stacking faults) or amorphous regions, generated by severe plastic deformation processes. Mechanical properties of the material are thus important to facilitate activation properties. For instance, brittle materials can be easily pulverized and activated because they expose fresh surfaces. The effect of substitution on modifying microstructure is thus important to be characterized and understood.

TiFe has poor absorption and desorption kinetic properties, however the partial substitution of Fe (e.g. with Ni, Mn, Cu) can improve the rate of the processes. This effect can be explained by the highly catalytic effect of Ni towards hydrogen chemisorption or, in the case of Mn, by the lowering of plateaus pressure, which increases the driven force for a given applied pressure. Kinetics can also be influenced by the type of synthesis (e.g. ball milling)[208] or thermal treatment, mainly due to microstructural modifications (nanostructuration, defects formation, etc.) that will not be discussed here in detail.

**Thermodynamics**

General correlations have been evidenced regarding geometric models stating that a linear relationship exists between the logarithm of the plateau pressure and the volume of the unit-cell or interstitial sites of the alloy. The larger the volume, the more stable the hydride is.[105,136,209]

Shinar *et al.* collected different PCI curves at different temperatures and the related thermodynamics of $TiFe_{0.80}M_{0.20}$ alloys substituted by $M$ = Mn, Cr, V, Co, Ni and Cu, evidencing that only the Mn-substituted hydrides showed clearly two flat plateaus, while, Cr- and V-substituted ones present two sloped plateaus and the formation of secondary phases.[210] For Co, Ni and Cu, only one plateau was observed.[210] Slopped plateaus can be caused by chemical inhomogeneity in substituted TiFe-alloys. Furthermore, size distribution of interstices caused by inhomogeneous distribution of substituents is claimed to cause sloping plateaus.[138,211] Another reason related to sloped plateaus is derived from continuous re-distribution of metallic elements during hydrogenation to form the fully hydrogenated compound, if fast diffusion and full local equilibrium (i.e. ortho-equilibrium) occur, instead of equilibrium limited to hydrogen as fast diffuser (i.e. para-equilibrium) .[212] The addition of a ternary substituent can reduce as well the hysteresis effect.[213]

Many different parameters should be considered and can be correlated to the substitution effects of different elements to hydrogenation properties of the TiFe intermetallic compound. As cited before, the logarithm of the first plateau pressure can be inversely correlated with the unit cell volume of the TiFe phase. During the structural transition from the intermetallic compound to the hydride, the lattice expansion/distortion is related to the thermodynamic stability of the hydrides and to the dimension of the octahedral interstices where hydrogen is hosted. Geometrical features can also explain why, in some cases, the formation of the $\gamma$ phase is hindered, together with electronic effects associated to the difference in valence electrons between the substituted element and Ti or Fe. Substitution does not only modify the volume of the interstices, but also the bond strength between hydrogen and metal atoms. The strength of *M*-H bonds determines the hydrogenation properties too. Yukawa *et al.* demonstrated that molecular orbital method and the study of electronic structure can clarify the relationship between bond order and strength, and their relation to hydrogen storage properties in substituted TiFe materials.[214] They reported that the hydrogen atoms interact stronger with Fe rather than with Ti, thus the chemical bond energy or the type of interaction with this atom or any substitutional elements modify the stability of the hydride phase.[214]

Moreover, substitution influences the hydrogenation properties of the material, modifying its Gibbs free energy. Thus, hydrogenation properties can be related to the entropy of mixing of hydrogen to enter the interstitial voids or the affinity of hydrogen with the substitutional elements.[215,216] This can explain the correlation between different elements and the variation of enthalpy or entropy of hydrides formation.[215,216] In 1981, Mintz *et al.* already discussed the influence of substitution in ternary TiFe alloys, reporting a sequence of hydride stability (from more to less stable Cr>Mn>Ni>Co>Fe).[138,213] The authors showed that a linear correlation of hydrogenation enthalpy as a function of substituent content is experimentally verified.[213]

In **Figure 7**, enthalpy and entropy values (**a** and **b,** respectively) for the first and second plateau in TiFe-type alloy collected in this review are reported as a function of values of the TiFe cell volume. Dashed lines for ΔH (39 kJ/mol) required at 25 °C to obtain 0.1 MPa plateau pressure with an expected entropy for gas-solid transition (130 J/molK) are reported as reference values.

As a rule, it can be observed that the hydrogenation enthalpy of both plateaus correlates linearly with the cell volume of the TiFe phase (**Figure 7, a**). However, some highly dispersed value of enthalpy can be found in the literature for the same alloy or same cell volume, as it can be observed at a fixed cell volume value in **Figure 7, a**. Value dispersion can be related to different methodologies, different conditions of PCI curves determination or material processing and experimental errors.

A linear trend of hydrogenation enthalpy related to the first plateau in absorption as a function of TiFe cell volume is evidenced in **Figure 7, a**. The same trend can be visualised for the first plateau desorption enthalpy values, which however are higher with respect to the first plateau absorption enthalpies. A clear trend for the second plateau enthalpies, both in absorption and desorption, cannot be visualized, even if a linear proportion as a function of TiFe cell volume could be suggested.

On the other hand, hydrogenation entropy does not significantly variate with cell volume (Figure 7, b). In fact, entropy change is mostly related to the gas to solid transition of hydrogen. Nevertheless, it must be underlined that even if entropy values are rather constant, they are higher for the second than for the first plateau, being both below the expected entropy change for the gas-solid transition (130 J/molK). Entropy values for the first plateau in absorption results lower than the one in desorption. Entropy values for the second plateau evidence higher values with respect to the first plateau, with higher values in desorption than absorption, as already reported for the first plateau.

**Cycling and resistance to poisoning**

Cycling and poisoning of the material can cause a drop or regular decrease in capacity, thus reducing the amount of hydrogen that can be stored. Two main causes have been argued for this fact. First, the formation of defects because of induced stress during hydriding and dehydriding reactions and, second, the formation of unreactive sites due to oxidation by low purity hydrogen, coming from commercial electrolyser mainly containing $H_2O$ as contaminant. In addition, intrinsic degradation on cycling could be caused by alloy disproportion into $TiH_2$ and Fe, which is driven by thermodynamics.[217] Moreover, hydrogenation capacity can be reduced on cycling by decomposition of TiFe into $TiO_2$ and Fe due to oxygen (or moisture) uptake. In this sense, the addition of a third element to the alloy can be beneficial if it acts as a getter element for oxidation and if TiFe can be easily recovered and oxides removed in mild conditions of annealing under hydrogen atmosphere or by leaching.

## Application of TiFe for hydrogen storage

A recent review from Lototskyy *et al.*[218] reports some examples of MH-tank and FC system developed at the lab-scale using metal hydrides. Most of systems are based on $AB_2$ ($(TiZr)(Mn,Fe,V)_2$-type) and $AB_5$ ($LaNi_5$-type) intermetallic compounds, which are usually preferred thanks to their low pressure and temperature working conditions, despite their moderate gravimetric capacity. In the following applications of TiFe-based alloys will be reported and reviewed.

Back in the ‘70s, the Brookhaven National Laboratory studied and developed bulk storage techniques for hydrogen using TiFe. The program consisted of a variety of activities which include engineering analysis and design of a large bulk hydrogen storage facility, engineering-scale tests, work on the selection and development of suitable iron-titanium alloys, and the construction of a large prototype energy storage system.[219] Strickland *et al.* implemented a small test bed of 38 kg of iron titanium hydride as storage media. The maximum hydrogen storage capacity, under the studied operating conditions, was 1.19 wt.% for the FeTi alloy and a uniform hydrogen flow rate of 9 normal litre per minute was sustained for a 10-hour transfer period.[220] Vessels of different dimensions were considered and tested for volumetric expansion with different alloy loading.[221] In the study, $TiFe_{0.70}Mn_{0.18}$ alloy was also considered.[221] Finally, the reservoir has been scaled up to store 6 kg hydrogen using 405 kg of TiFe.[222] This latter prototype system was built for the Public Service Electric and Gas Company of New Jersey to study the feasibility of storing off-peak electrical energy through the use of a water electrolyser, a hydride reservoir and a fuel cell stack.[222]

The use of TiFe in hydride beds has been developed for storing and supplying hydrogen fuel in power plant and automotive applications.[223] An hydride bed was built at Brookhaven National Laboratory for the Public Service Electric and Gas Co., with a release rate roughly constant of approximately 0.036 Ib $ft^{-1}$ h (0.000015 Pa s).[222] Finally, a techno-economic assessment was performed for a hydrogen-chlorine energy storage system for electric utility load levelling and peak shaving applications that involved hydrogen storage in TiFe-alloy.[224]

Recently, a conventional bench-scale tank based on TiFe carriers have been developed in Japan.[197] Endo *et al.* developed a TiFe-based material to be used in an integrated system with an electrolyser and a fuel cell. The material is a modified $TiFe_{0.80}Mn_{0.20}$ alloy, with the introduction of some V and annealed at 1100 °C for 24 h.[197] The activation of the material was performed at modest temperature and low pressure, according to Japanese safety regulations (i.e. approx. 80 °C and 2 bar).

Vacuum condition at the bench scale are crucial for decreasing incubation time during activation.[197] Furthermore, a similar material was later used in a recent study for a hydrogen energy utilization system in a zero emission building.[225] The integrated system used 520 kg of *M*H for 80 $Nm^3$ of total hydrogen storage and it has been tested under different weather conditions.[226]

They later on demonstrate that limited use of auxiliaries for cooling can be implemented reducing power consumption in a bench-scale hydrogen system (Hydro Q-BiC), reaching an overall energy efficiency of 60% (power-to-hydrogen-to-power/heat).[227]

In Europe, the only demonstrative energy storage system based on TiFe carrier is under development in South Tirol by GKN. It will demonstrate the use of hydrogen carrier for about 10 kg $H_2$, working close to room temperature and pressure.

Finally, the European HyCARE project is willing to develop an integrated system for stationary storage of hydrogen in TiFe-based material efficiently coupled with a electrolyser, fuel cell and heat storage system based on phase change material.[18,19] The total amount of hydrogen that will be stored is expected to be 50 kg, in approx. 4 tons of TiFe-based carrier.

## Conclusions

As a conclusion, TiFe remains an attractive alloy for hydrogen storage regarding its low cost and significant capacity. However, it is worth noting that the binary compound has some drawbacks to be overcome by suitable substitutions, strongly improving activation and kinetics. However, thermodynamics and hydrogenation properties are also influenced, and often a drop in capacity can be observed. This review evidences that Mn is a key element as substituent in TiFe system. In addition, V, and the synergic effect of Mn and V can be exploited. However, it should be kept in mind that V is a CRMs for Europe. On the other hand, many other elements such as Mg, Ta, Zr, Cr, Co, Ni, Cu and S are not suitable for industrial upscaled materials. Zr, Cr and S lead to sensible reduction of storage capacity. Mg, Ta, Co, Ni are CRMs. Cu increases sensibly the pressure gap between the first and second plateau. Effect of contaminants during synthesis (such as B, C, N, O) should be studied and understood with care because they can negatively influence hydrogenation properties of the material. As a general assertion, bi- and multi-substituted alloys can combine some synergic and beneficial effects, which however are not drastically better than for mono-substituted cases.

The study performed in this work evidenced that systematic studies on TiFe-substituted alloys are scarce in the literature. In the last years, a high number of reports have been published with only partial characterisation of the studied alloys, making hard the understanding of the full picture concerning activation, thermodynamics, kinetics and chemistry of the studied systems. To further understand the role of elemental substitution in TiFe on the modification of hydride stability and hydrogenation properties, there is a strong need of complete studies including compositional, structural, microstructural, activation, kinetic, and thermodynamic data. More attention should be focused on the determination of elemental substitution in TiFe, perhaps by electron microprobe analysis (EMPA) or by coupling scanning electron microscope (SEM) and energy-dispersive x-ray spectroscopy (EDX) mapping to analytically determine the alloy (TiFe-phase) exact composition. Structural studies should also include experimental or computational evidences of possible elemental substitution in Ti or Fe position. The determination of cell parameter of TiFe, and relative abundances of secondary phases determined by microscopy and X-ray diffraction (coupled to Rietveld analysis) methods are useful information to understand the geometrical implication of substitution. Neutron studies are complementary for the full solution of crystal structure of substituted systems. The review was not focused on neutron studies on TiFe systems, but there is a big lack of these determinations. Some studies can be found in the following references.[61–63,66,67,69,70,74,228–231]

The study and assessment of ternary and higher phase diagrams can also help in the definition of phase boundaries, thus elucidating the formation of secondary phases and their possible role in enhancing activation properties. Mechanical properties and microstructural studies of these materials are also of great interest because they can be related to enhanced activation process and easy handling during industrial processing such as crushing process or pellet preparation.

In conclusion, the determination of thermodynamics in these systems must be conducted with care. Annealed samples can be representative of equilibrium phases, and the study of PCI curves with sufficient waiting time for the determination of equilibrium pressure points both in absorption and desorption is essential. PCI curves should also be determined in an appropriately large temperature range to guarantee a correct determination of enthalpy and entropy by the Van't Hoff plot. Coupling calorimetric determinations can confirm and complete the thermodynamic study. Cycling properties and resistance to poisoning are not frequent in literature studies, but they are of great interest for real applications of these materials. More efforts should focus on these studies and the definition of the mechanisms involved, in order to solve dropping capacity problems in case of not pure hydrogen produced by commercial electrolysers. Few other properties, which were not included in this review, are also of great interest for real applications. They include thermal properties (thermal conductivity, heat capacity), density and porosity determination. The definition of these parameters will help the upscale and development of prototype tanks for solid-state hydrogen storage based on TiFe systems.

## Conflicts of interest

There are no conflicts to declare.

## Acknowledgement

The project leading to this publication has received funding from the Fuel Cells and Hydrogen 2 Joint Undertaking (JU) under grant agreement No 826352, HyCARE project. The JU receives support from the European Union's Horizon 2020 research, Hydrogen Europe, Hydrogen Europe Research, innovation programme and Italy, France, Germany, Norway, which are all thankfully acknowledged.

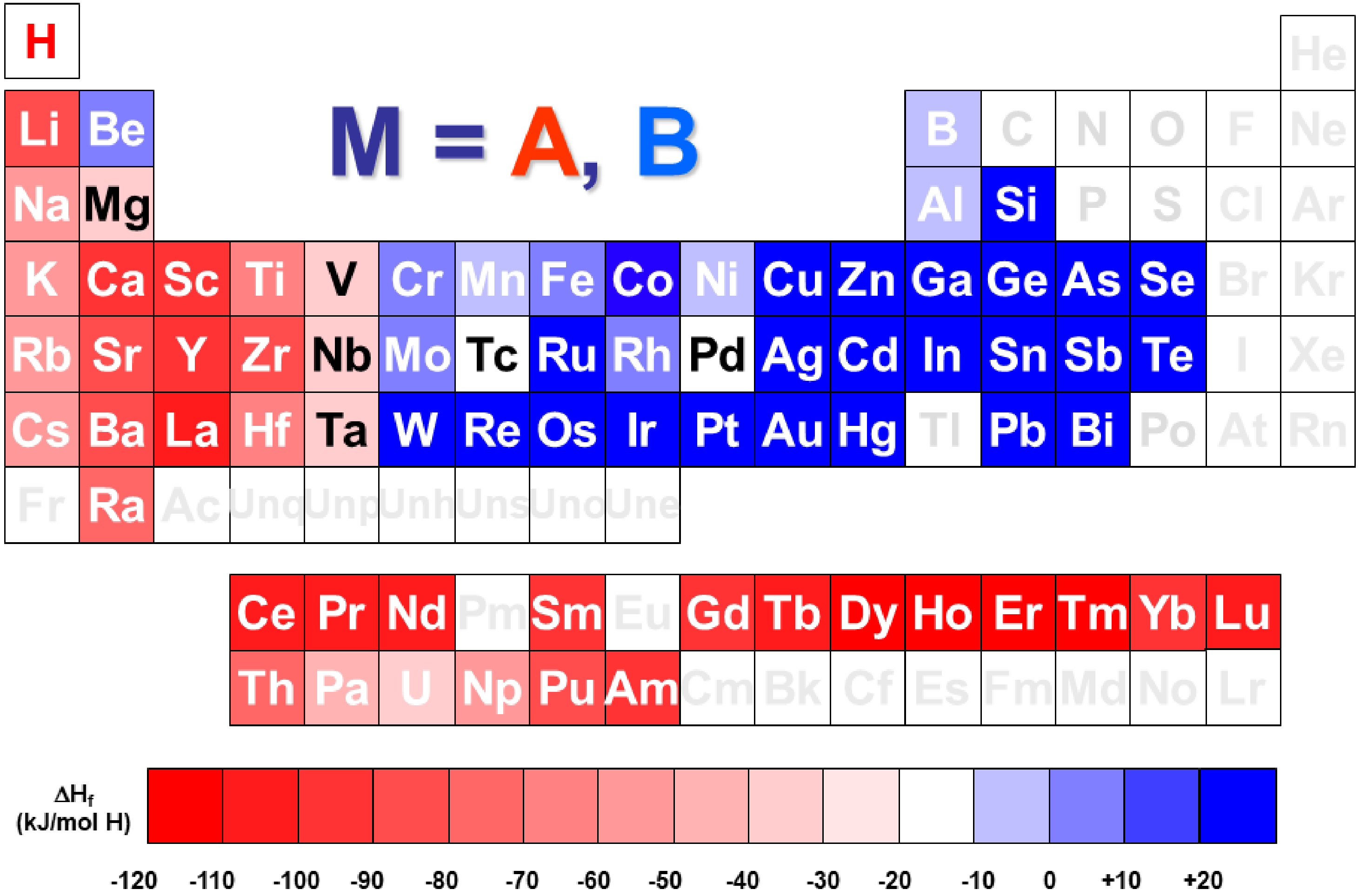

**Figure 1** – Periodic table of the elements showing the formation enthalpy of binary *M*-H Metal hydrides, and the relative classification of A (in red) or B-type (in blue) *M* elements. Formation enthalpy was chosen from literature values.[60,205,206,232,233]

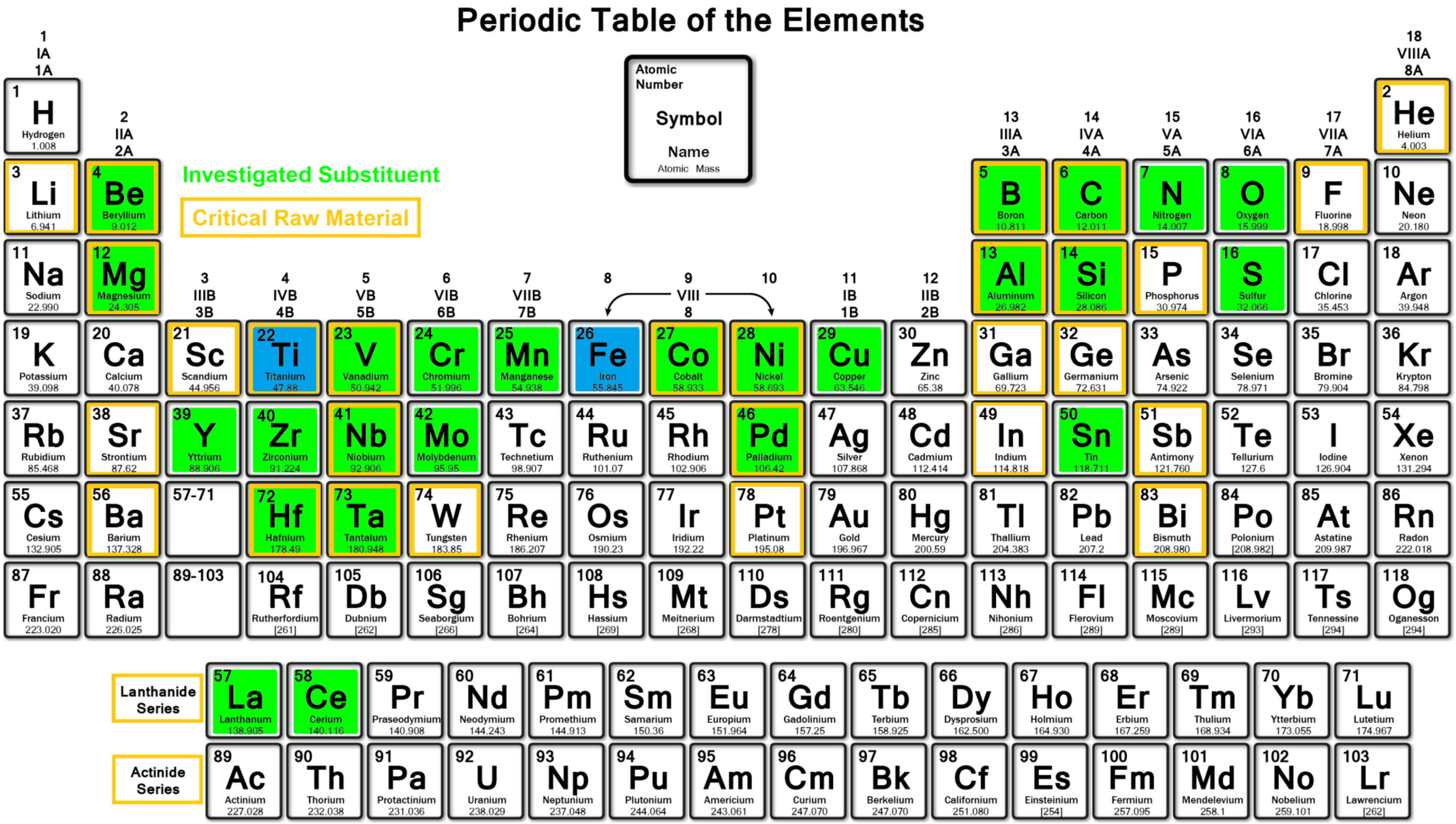

**Figure 2** – Periodic table showing the investigated substituting elements (highlighted in green) in TiFe (highlighted in blue). Orange frame highlight the critical raw materials for Europe.[21]

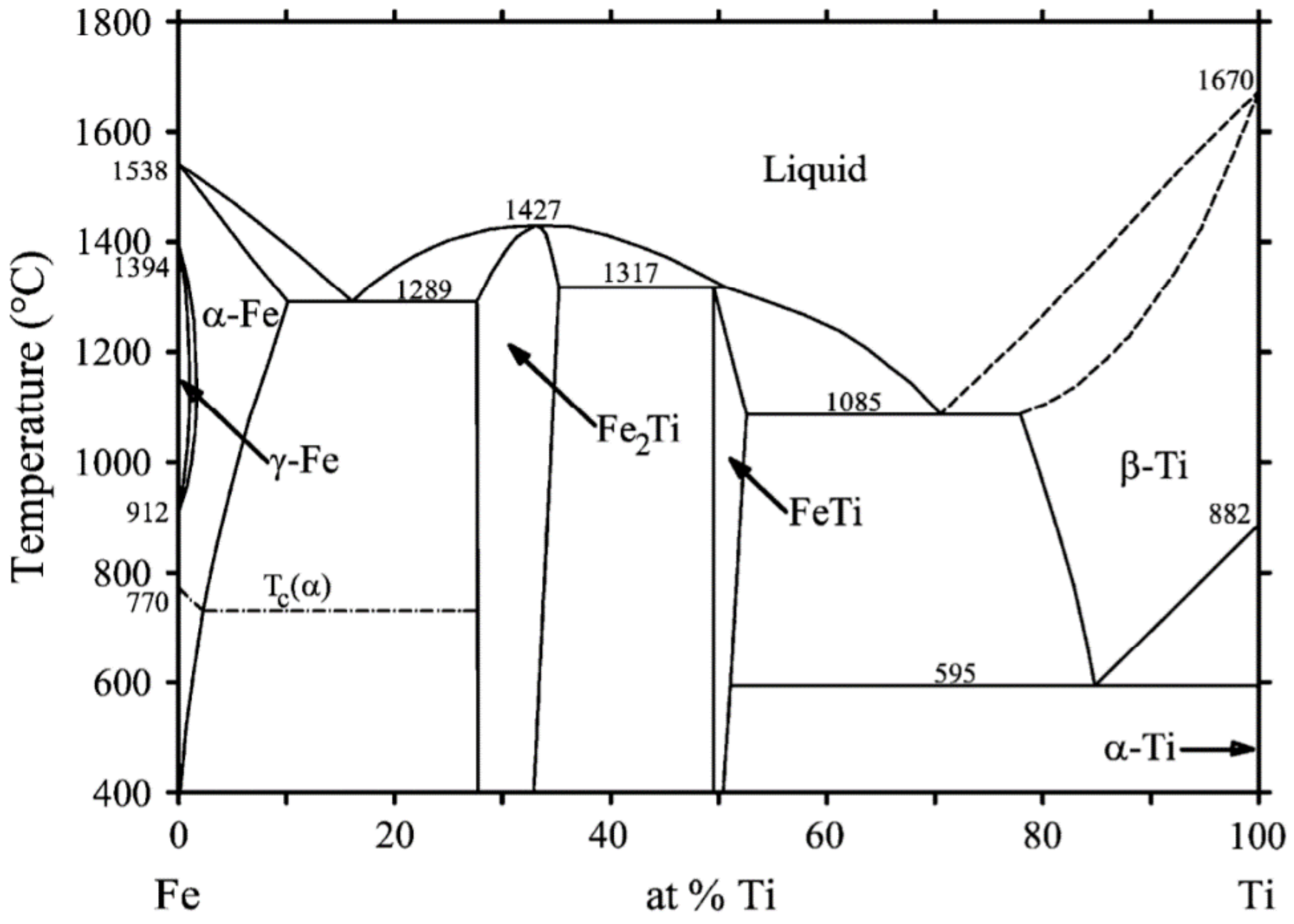

**Figure 3 -** Fe-Ti phase diagram. Reproduced with permission from [25,26].

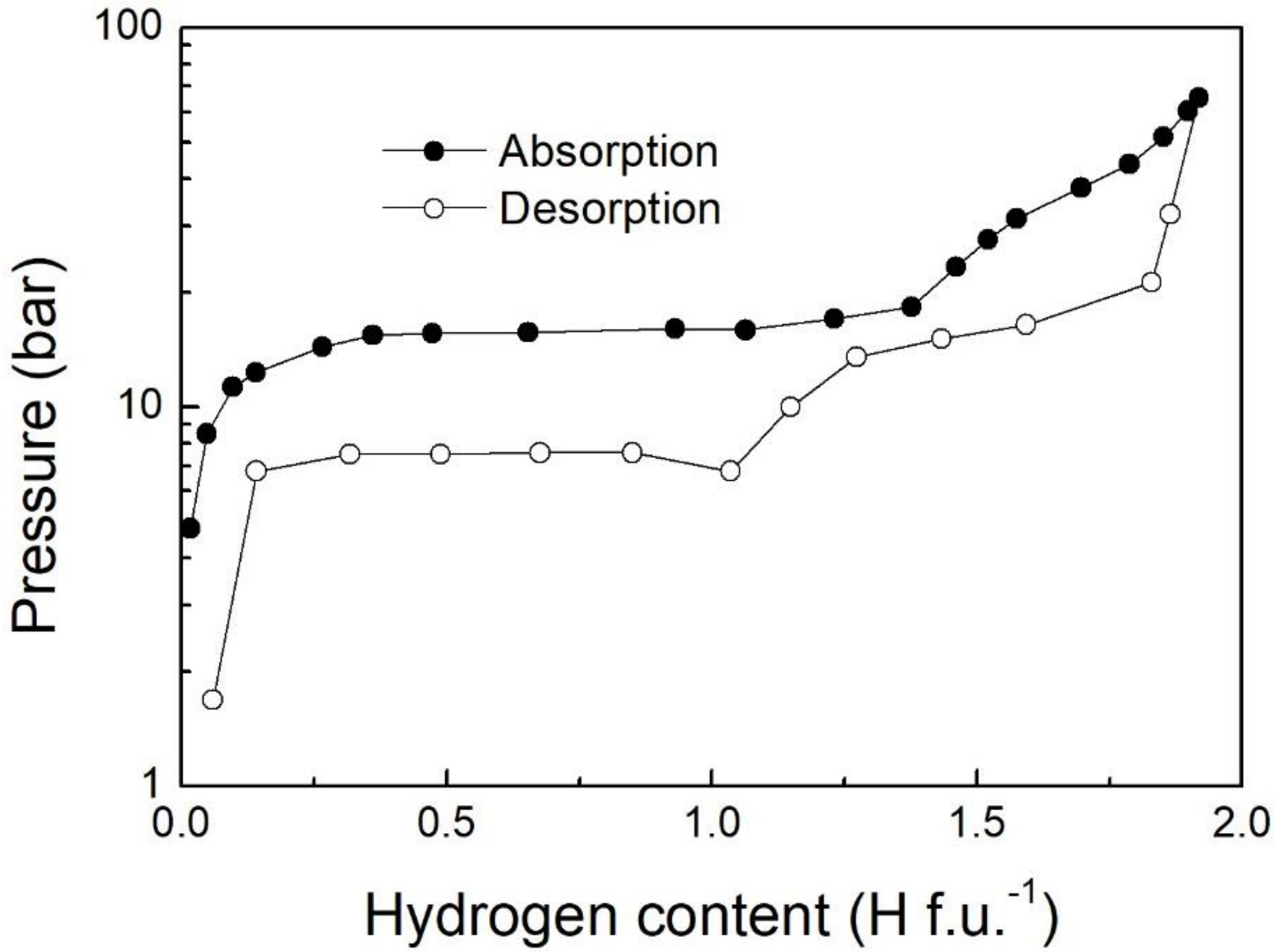

**Figure 4 -** Pressure-Composition-Isotherm of TiFe performed at 40 °C.

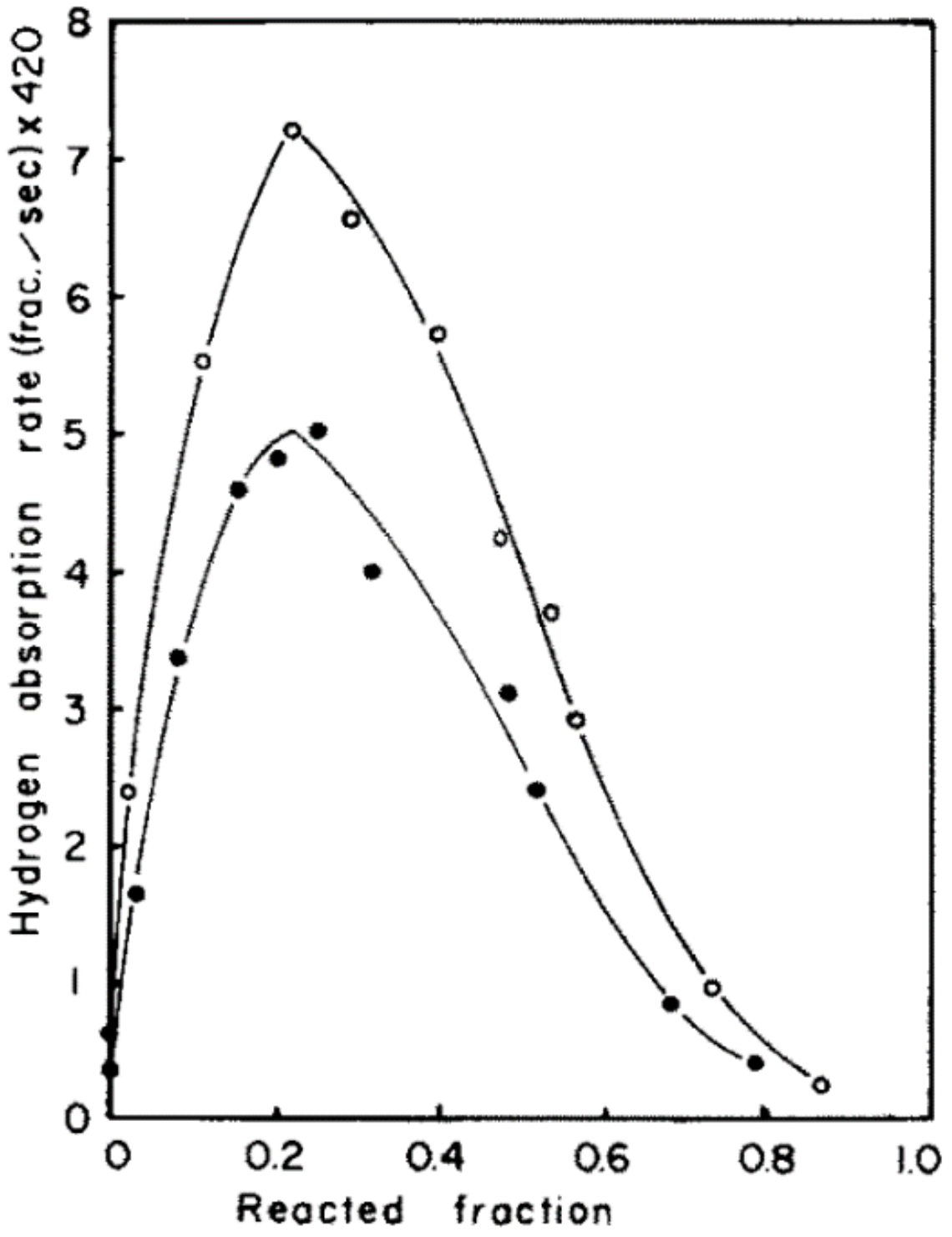

**Figure 5 -** Hydrogen absorption rate as a function of reacted fraction at 20 °C. The curve with black dots was obtain under 20 bar of hydrogen, whereas the one with white dots under 24 bar. Reproduced with permission from [94].

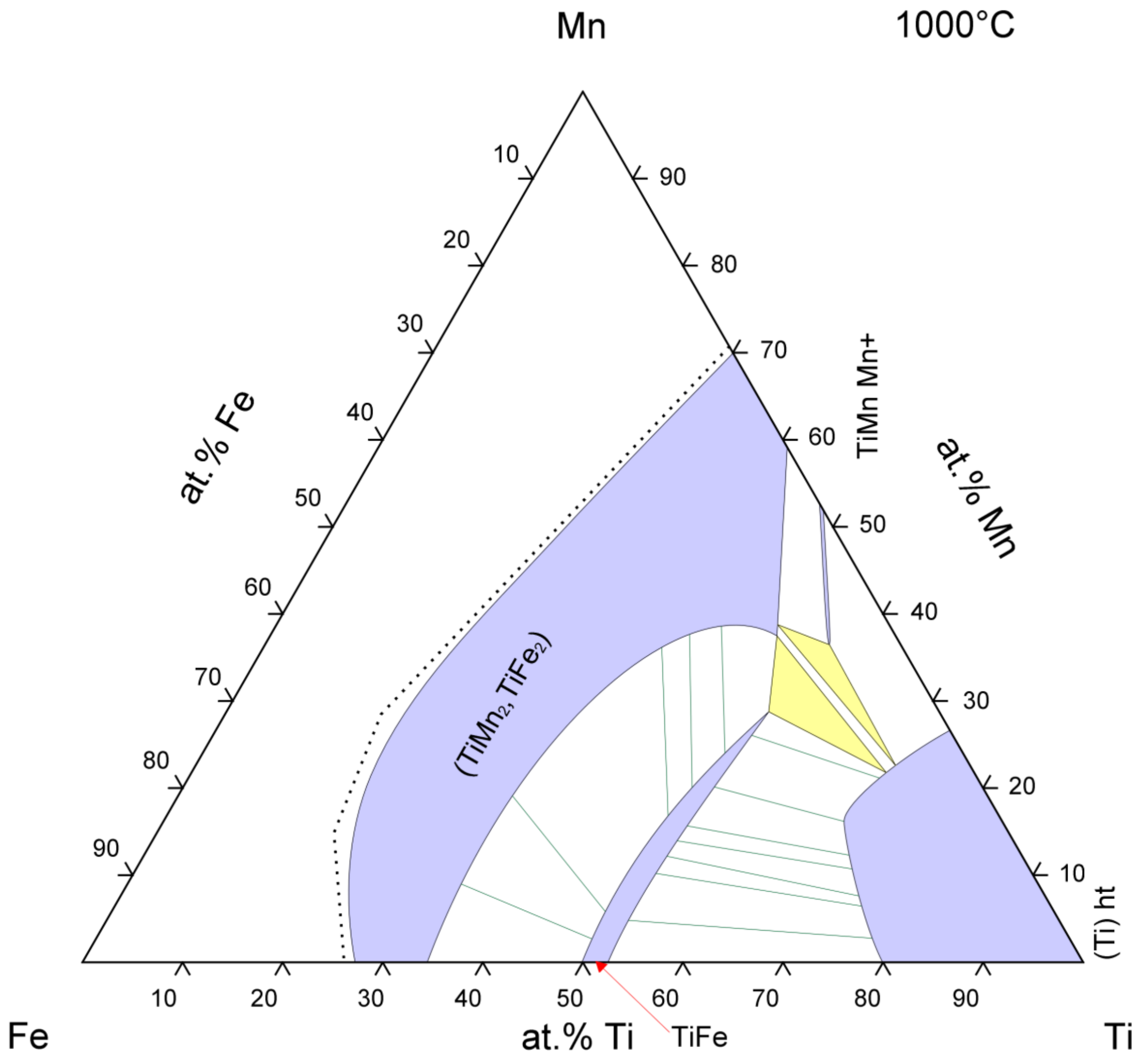

**Figure 6** – Isotherm section at 1000 °C of the Ti-Fe-Mn phase diagram. Reproduced with permission from [22].

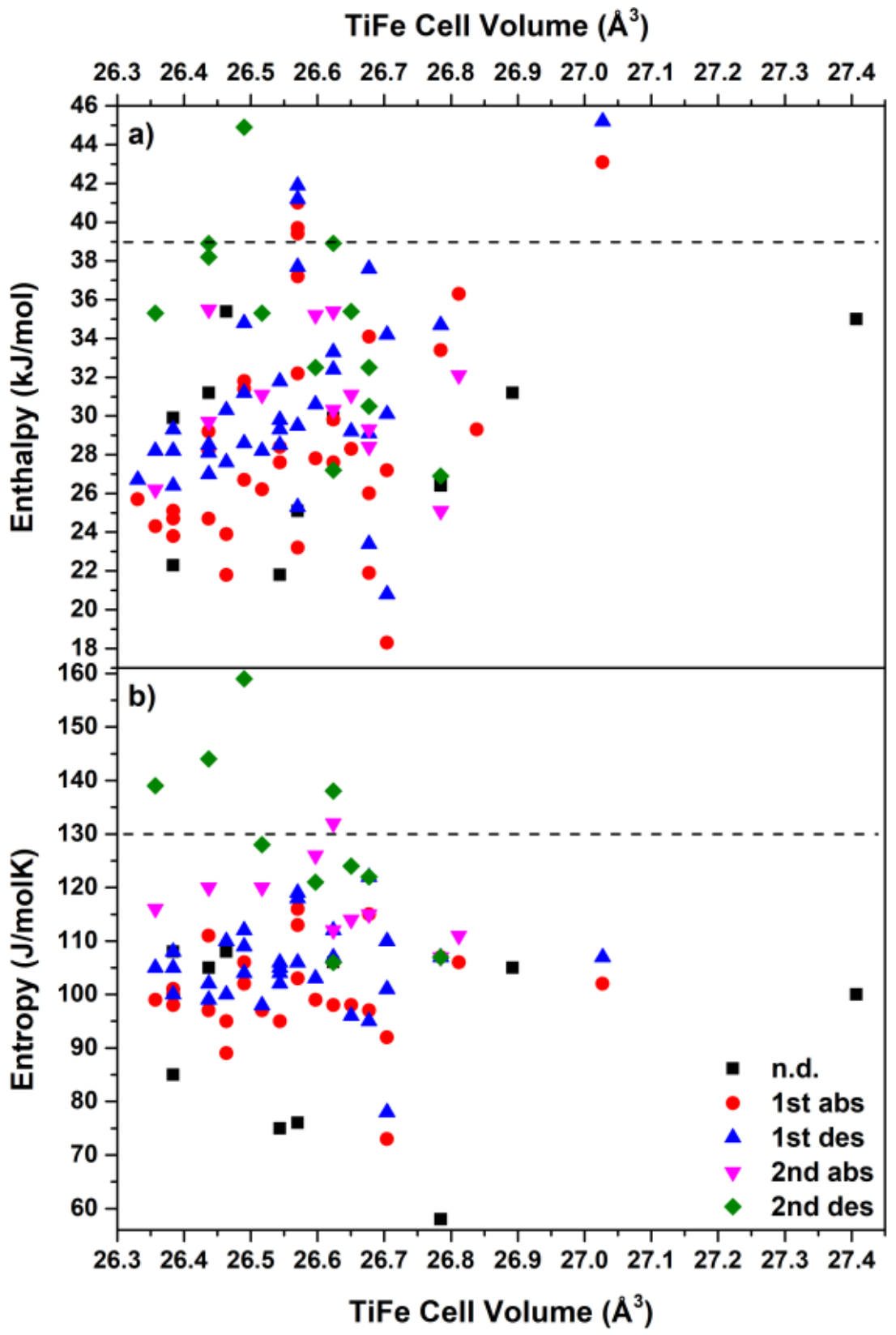

**Figure 7 –** Correlation between cell volumes of substituted TiFe phase with absolute a) enthalpy and b) entropy values of hydrogenation. Black dashed lines for ΔH (39 kJ/mol) required at 25°C to obtain 0.1 MPa plateau pressure with expected entropy for gas-solid transition (130 J/molK). Values related to the first plateau in absorption are reported as red circle, blue triangle (up) for first plateau values in desorption, pink triangle (down) for second plateau values in absorption, green diamond for second plateau values in desorption, and black square when the plateau was not defined.

**Table 1 –** Thermodynamic properties of TiFe during hydrogen absorption/desorption obtained from Van't Hoff equation and direct calorimetric measurements.

| ***Plateau*** | ***-ΔH<sub>abs</sub>*** kJ mol$^{-1}$ | ***-ΔS<sub>abs</sub>*** J mol$^{-1}$ K$^{-1}$ | ***ΔH<sub>des</sub>*** kJ mol$^{-1}$ | ***ΔS<sub>des</sub>*** J mol$^{-1}$ K$^{-1}$ | ***T*** °C | ***References and notes*** ***Technique: van't Hoff*** |
|---|---|---|---|---|---|---|
| First | | | 28.1 | 106 | | [68], Prepared from zone-refined melting of Ti and Fe in an arc furnace |
| Second | | | 33.7 | 132 | | |
| First | 25.4 | 104 | 25.6 | 97 | | [93], Prepared by induction melting of Ti and Fe |
| Second | 33.2 | 137 | 31.6 | 125 | | |
| First | 26.4 | 113 | 27.8 | 107 | | [54], Nanocrystalline TiFe prepared by mechanical alloying of Ti and Fe powders |
| Second | | | | | | |
| First | | | 31.0 | 118 | | [66], Prepared by arc melting of Ti and Fe and loaded with Deuterium |
| Second | | | | | | |
| First | 24.3 | 100 | 27.4 | 103 | | [178], Prepared by induction melting of Ti and Fe, annealed at 1000 °C for 1 week |
| Second | | | | | | |
| | | | | | | ***Technique : Calorimetry*** |
| First | 24.6 | 92 | 24.8 | 85 | 1 | [93], Prepared by induction melting of Ti and Fe |
| Second | 29.4 | 114 | 30.6 | 111 | 1 | |
| First | 24.2 | 99 | 24.2 | 90 | 25 | |
| Second | 29.8 | 128 | 32.0 | 126 | 25 | |
| First | 23.4 | 100 | 23.4 | 92 | 41 | |
| Second | 26.6 | 122 | 27.8 | 117 | 41 | |
| First | 22.8 | 104 | 23.4 | 98 | 71 | |
| Second | 28.4 | 134 | 28.0 | 126 | 71 | |
| First | 24.9 | | 27.4 | | 35 | [43], Prepared by mechanical alloying of Ti and Fe powders |
| Second | | | | | 35 | |
| First | 23.0 | | 27.2 | | 35 | [43], As received commercial TiFe powder |
| Second | 33.9 | | 35.2 | | 35 | |

**Table 2** – Alloy composition, secondary phases, TiFe-phase lattice parameter and main hydrogenation properties of reviewed TiFe-*M* systems.
$^{a}$: absorption, $^{d}$: desorption, $_{1st}$: 1$^{st}$ plateau, $_{2nd}$: 2$^{nd}$ plateau, RT: room temperature.

| TiFe-M | Ti %at | Fe %at | M %at | Secondary Phases | TiFe a (Å) | Capacity (%wt) Conditions | PCI Thermodynamics | Ref. |
|---|---|---|---|---|---|---|---|---|
| **-Mg** | | | | | | | | |
| $TiFeMg_{0.04}$ | 49.0 | 49.0 | 2.0 | | 2.982 | 1.10<br>22 °C, 20 bar | 22 °C | [108] |
| $TiFeMg_{0.08}$ | 48.0 | 48.0 | 4.0 | α-Fe | 2.990 | | | [108] |
| $TiFeMg_{0.13}$ | 47.0 | 47.0 | 6.0 | α-Fe | 2.998 | | | [108] |
| **-Be** | | | | | | | | |
| $TiFe_{0.90}Be_{0.10}$ | 50.0 | 45.0 | 5.0 | | 2.977 | 1.05<br>50 °C, 10 bar | 21,$^{d}$ 50, 80 °C<br>ΔH = 29.9 kJ/mol<br>ΔS = 108 J/molK | [110] |
| $TiFe_{0.80}Be_{0.20}$ | 50.0 | 40.0 | 10.0 | | 2.977 | | 21,$^{d}$ 50, 80, 125$^{d}$ °C | [109] |
| | | | | | 2.979 | 1.00<br>50 °C, 10 bar | 21,$^{d}$ 50, 80, 125$^{d}$ °C<br>ΔH = 31.2 kJ/mol<br>ΔS = 105 J/molK | [110] |
| $TiFe_{0.70}Be_{0.30}$ | 50.0 | 35.0 | 15.0 | | 2.980 | 0.95<br>50 °C, 10 bar | 50, 80, 125$^{d}$ °C<br>ΔH = 35.4 kJ/mol<br>ΔS = 108 J/molK | [110] |
| **-Zr** | | | | | | | | |
| $TiFeZr_{0.04}$ | 49.0 | 49.0 | 2.0 | | 2.979 | 1.10<br>22°C, 37 bar | 22 °C | [117] |
| $TiFeZr_{0.04}$ | 48.9 | 48.9 | 2.2 | $TiFe_2$ | | 1.60<br>40°C, 20 bar | 40 °C | [119] |
| | | | | ($Ti_2Fe$) | | 1.60<br>RT, 45 bar | | [111] |
| | | | | ($Ti_2Fe$, $TiFe_2$) | 2.983 | | | [115,116] |
| | | | | ($TiFe_2$) | 2.980 | | | [120] |
| $TiFeZr_{0.05}$ | 48.8 | 48.8 | 2.4 | | 2.983 | | RT | [114] |
| $TiFeZr_{0.08}$ | 48.0 | 48.0 | 4.0 | | 2.995 | | | [117] |
| $TiFeZr_{0.09}$ | 47.9 | 47.8 | 4.3 | ($Ti_2Fe$, $TiFe_2$) | 2.986 | | | [116] |
| $TiFeZr_{0.10}$ | 47.7 | 47.7 | 4.6 | ($Ti_2Fe$, $TiFe_2$) | 2.980 | | | [115] |
| $TiFeZr_{0.13}$ | 47.0 | 47.0 | 6.0 | | 2.995 | | | [117] |
| $TiFeZr_{0.16}$ | 46.4 | 46.4 | 7.2 | ($Ti_2Fe$, $TiFe_2$) | 2.984 | | | [115] |
| $TiFeZr_{0.17}$ | 46.0 | 46.0 | 8.0 | Fe | 2.997 | | | [117] |
| $TiFeZr_{0.22}$ | 45.1 | 45.1 | 9.8 | ($Ti_2Fe$, $TiFe_2$) | 2.981 | | | [115] |
| $TiFe_{0.95}Zr_{0.05}$ | 50.0 | 47.5 | 2.5 | | 2.980 | | RT | [114] |
| $Ti_{1.3}Fe_{0.80}Zr_{0.20}$ | 56.5 | 34.8 | 8.7 | $Ti_2Fe$,<br>$TiFe_2$, Ti | | 1.20<br>200 °C, 20 bar | 100 °C | [112] |
| $Ti_{0.99}FeZr_{0.01}$ | 49.5 | 50.0 | 0.5 | | | | 30 °C [d] | [118] |
| $Ti_{0.95}FeZr_{0.05}$ | 47.5 | 50.0 | 2.5 | | 2.985 | | RT | [114] |
| $Ti_{0.90}FeZr_{0.10}$ | 45.0 | 50.0 | 5.0 | | | | 20, 40 °C [d]<br>$-\Delta H_{1st}{}^{a}$ = 21.8 kJ/mol | [113] |
| | | | | | | | 30 [d], 45, 60 °C<br>$\Delta H^d$ = 28.9 kJ/mol<br>$\Delta S^d$ = 105 J/molK<br>$-\Delta H^a$ = 26.2 kJ/mol<br>$-\Delta S^a$ = 102 J/molK | [118] |
| $Ti_{0.84}FeZr_{0.16}$ | 42.0 | 50.0 | 8.0 | | | | 50, 100, 150, 200 °C [a] | [234] |
| $Ti_{0.80}FeZr_{0.20}$ | 40.0 | 50.0 | 10.0 | | | | 40 °C [d] | [113] |
| | | | | | | | 30 °C [d] | [118] |
| $Ti_{0.50}FeZr_{0.50}$ | 25.0 | 50.0 | 25.0 | | | | 40 °C [d] | [113] |
| **-Hf** | | | | | | | | |
| $TiFeHf_{0.02}$ | 49.4 | 49.4 | 1.2 | $TiFe_2$ | 2.982 | | RT | [121] |
| $TiFeHf_{0.04}$ | 48.9 | 48.9 | 2.2 | $TiFe_2$ | 2.986 | | RT | [121] |
| $TiFeHf_{0.07}$ | 48.3 | 48.3 | 3.4 | $TiFe_2$ | 2.989 | | RT | [121] |
| $TiFeHf_{0.09}$ | 47.8 | 47.8 | 4.4 | $TiFe_2$ | 2.990 | | RT | [121] |
| **-V** | | | | | | | | |
| $Ti_{0.95}FeV_{0.06}$ | 47.1 | 49.8 | 3.1 | | | 1.10<br>54 °C, 20 bar | | [126] |
| $Ti_{0.94}FeV_{0.06}$ | 47.0 | 50.0 | 3.0 | $TiFe_2$, $Ti_4Fe_2O$ | 2.976 | | 30 °C [d] | [104] |
| $Ti_{0.96}FeV_{0.04}$ | 48.0 | 50.0 | 2.0 | | | | 25 °C [d] | [235] |
| $Ti_{0.90}FeV_{0.10}$ | 45.0 | 50.0 | 5.0 | $TiFe_2$, $Ti_4Fe_2O$ | 2.976 | | 30 °C [d] | [104] |
| $Ti_{0.80}FeV_{0.20}$ | 40.0 | 50.0 | 10.0 | $TiFe_2$ | 2.971 | | | [125] |
| $Ti_{0.96}Fe_{0.94}V_{0.10}$ | 48.0 | 47.0 | 5.0 | | | | 20, 40, 60 °C [d] | [124] |
| $Ti_{0.94}Fe_{0.96}V_{0.10}$ | 47.0 | 48.0 | 5.0 | $Ti_4Fe_2O$ | 2.980 | | 30 °C [d] | [104] |
| | | | | | 2.979 | | 25, 65, 100 °C<br>$\Delta H^d$ = 27.0 kJ/mol<br>$\Delta S^d$ = 99 J/molK<br>$-\Delta H^a$ = 29.2 kJ/mol<br>$-\Delta S^a$ = 111 J/molK | [125] |
| $Ti_{0.96}Fe_{0.68}V_{0.36}$ | 48.0 | 34.0 | 18.0 | Bcc, $TiFe_2$ | 3.000 | | | [125] |
| $TiFe_{0.90}V_{0.05}$ | 51.3 | 46.2 | 2.6 | $TiFe_2$, β-Ti | 2.987 | 1.96<br>25°C, 25 bar | 25 °C | [236] |
| $TiFe_{0.94}V_{0.06}$ | 50.0 | 47.0 | 3.0 | β-Ti, $Ti_4Fe_2O$ | 2.989 | | 30 °C [d] | [104] |
| $TiFe_{0.90}V_{0.10}$ | 50.0 | 45.0 | 5.0 | $TiFe_2$, β-Ti | 2.989 | 1.68<br>25°C, 25 bar | 25 °C | [236] |
| | | | | β-Ti, $Ti_4Fe_2O$ | 2.995 | | 30 °C [d] | [104] |
| $TiFe_{0.80}V_{0.20}$ | 50.0 | 40.0 | 10.0 | | | | 54, 79, 102°C [d] | [210] |
| $TiFe_{0.50}V_{0.17}$ | 60.0 | 30.0 | 10.0 | Bcc | 2.999 | | | [125] |
| **-Nb** | | | | | | | | |
| $TiFeNb_{0.04}$ | 49.0 | 49.0 | 2.0 | | 2.979 | | 35 °C<br>$-\Delta H_{1st}{}^{a}$ = 28.2 kJ/mol<br>$\Delta H_{1st}{}^{d}$ = 28.5 kJ/mol<br>$-\Delta H_{2nd}{}^{a}$ = 35.5 kJ/mol<br>$\Delta H_{2nd}{}^{d}$ = 38.2 kJ/mol | [129] |
| | | | | α-Fe, $TiFe_2$ | 2.984 | | 35 °C<br>$-\Delta H_{1st}{}^{a}$ = 23.2 kJ/mol<br>$\Delta H_{1st}{}^{d}$ = 25.3 kJ/mol | [129] |
| $TiFeNb_{0.08}$ | 48.0 | 48.0 | 4.0 | α-Fe | | | | [129] |
| $TiFe_{0.90}Nb_{0.10}$ | 50.0 | 45.0 | 5.0 | β-Ti | 2.989 | | | [87] |
| $Ti_{0.96}FeNb_{0.04}$ + 1.0 wt.% $Fe_2O_3$ | | | | $TiFe_2$,<br>$Ti_{10}Fe_7O_3$ | | | 30 °C | [127] |
| $Ti_{0.92}FeNb_{0.08}$ + 1.0 wt.% $Fe_2O_3$ | | | | $TiFe_2$,<br>$Ti_{10}Fe_7O_3$ | | | 30 °C | [127] |
| $Ti_{0.90}FeNb_{0.10}$ + 1.0 wt.% $Fe_2O_3$ | | | | $TiFe_2$,<br>$Ti_{10}Fe_7O_3$ | | | 30 °C | [127] |
| $Ti_{0.88}FeNb_{0.12}$ + 1.0 wt.% $Fe_2O_3$ | | | | $TiFe_2$,<br>$Ti_{10}Fe_7O_3$ | | | 30 °C | [127] |
| **-Ta** | | | | | | | | |
| $TiFe_{0.90}Ta_{0.10}$ | 50.0 | 45.0 | 5.0 | | 2.990 | | 40 °C | [130] |
| $TiFe_{0.80}Ta_{0.20}$ | 50.0 | 40.0 | 10.0 | | 2.994 | | 40 °C | [130] |
| **-Cr** | | | | | | | | |
| $TiFeCr_{0.04}$ | 49.0 | 49.0 | 2.0 | | 2.973 | | | [134] |
| $TiFeCr_{0.08}$ | 48.0 | 48.0 | 4.0 | | 2.974 | 1.00<br>22 °C, 60 bar | 22 °C | [134] |
| $TiFeCr_{0.13}$ | 47.0 | 47.0 | 6.0 | | 2.972 | | | [134] |
| $TiFeCr_{0.17}$ | 46.0 | 46.0 | 8.0 | α-Fe | 2.972 | | | [134] |
| $TiFe_{0.95}Cr_{0.05}$ | 50.0 | 47.5 | 2.5 | $TiCr_2$ | | | 50 °C [d] | [131] |
| | | | | | | | 40 °C [d] | [136] |
| $TiFe_{0.90}Cr_{0.10}$ | 50.0 | 45.0 | 5.0 | | | | $\Delta H^{d}$ = 30.1 kJ/mol<br>$\Delta S^{d}$ = 101 J/molK | [213] |
| | | | | $TiCr_2$ | | | 50 °C [d] | [131] |
| | | | | | | | 40 °C [d] | [136] |
| | | | | Ti | 2.989 | | 10, 30, 50 °C<br>$\Delta H^{d}$ = 34.2 kJ/mol<br>$\Delta S^{d}$ = 110 J/molK<br>$-\Delta H^{a}$ = 27.2 kJ/mol<br>$-\Delta S^{a}$ = 92 J/molK | [132] |
| $TiFe_{0.80}Cr_{0.20}$ | 50.0 | 40.0 | 10.0 | | | | $\Delta H^{d}$ = 35.6 kJ/mol | [213] |
| | | | | | | | $\Delta S^{d}$ = 108 J/molK | |
| | | | | | | | $-\Delta H^{a}$ = 33.1 kJ/mol | |
| | | | | | | | $-\Delta S^{a}$ = 103 J/molK | |
| | | | | | | | 24, 40, 60 °C $^{d}$ | [210] |
| $TiFe_{0.70}Cr_{0.16}$ | 53.8 | 37.6 | 8.6 | $TiCr_2$ | 2.993 | 1.80 | | [141] |
| | | | | | | 25°C, 20 bar | | |
| **-Mo** | | | | | | | | |
| $TiFe_{0.90}Mo_{0.10}$ | 50.0 | 45.0 | 5.0 | | | | 40 °C $^{d}$ | [136] |
| | | | | β-Ti | 2.989 | | | [87] |
| | | | | | 2.989 | | 40 °C | [130] |
| $TiFe_{0.80}Mo_{0.20}$ | 50.0 | 40.0 | 10.0 | | 2.992 | | 40 °C | [130] |
| **-Co** | | | | | | | | |
| $TiFeCo_{0.04}$ | 49.0 | 49.0 | 2.0 | $TiFe_2$ | 2.988 | | 35 °C | [129] |
| | | | | | | | $-\Delta H_{1st}{}^{a}$ = 26.0 kJ/mol | |
| | | | | | | | $\Delta H_{1st}{}^{d}$ = 29.1 kJ/mol | |
| | | | | | | | $-\Delta H_{2nd}{}^{a}$ = 28.4 kJ/mol | |
| | | | | | | | $\Delta H_{2nd}{}^{d}$ = 30.5 kJ/mol | |
| | | | | α-Fe, $TiFe_2$ | 2.975 | | 35 °C | [129] |
| | | | | | | | $-\Delta H_{1st}{}^{a}$ = 25.7 kJ/mol | |
| | | | | | | | $\Delta H_{1st}{}^{d}$ = 26.7 kJ/mol | |
| $TiFeCo_{0.08}$ | 48.0 | 48.0 | 4.0 | α-Fe | | | | [129] |
| $TiFe_{0.90}Co_{0.10}$ | 50.0 | 45.0 | 5.0 | | | | $\Delta H^{d}$ = 30.6 kJ/mol | [213] |
| | | | | | | | $\Delta S^{d}$ = 106 J/molK | |
| | | | | | | | 40 °C $^{d}$ | [136] |
| | | | | | | | 20, 30, 40, 50 °C | [138] |
| $TiFe_{0.80}Co_{0.20}$ | 50.0 | 40.0 | 10.0 | | | | $\Delta H^{d}$ = 32.7 kJ/mol | [213] |
| | | | | | | | $\Delta S^{d}$ = 109 J/molK | |
| | | | | | | | 52.5, 82, 110 °C $^{d}$ | [210] |
| | | | | | | | $-\Delta H^{a}$ = 31.4 kJ/mol | |
| | | | | | | | $-\Delta S^{a}$ = 102 J/molK | |
| $TiFe_{0.50}Co_{0.50}$ | 50.0 | 25.0 | 25.0 | | | 1.10 | 80, 100, 120 °C $^{d}$ | [137] |
| | | | | | | RT, 30 bar | $\Delta H^{d}$ = 42.3 kJ/mol | |
| | | | | | | | $\Delta S^{d}$ = 123 J/molK | |
| | | | | | | | 80, 100, 120 °C | [139] |
| | | | | | | | $\Delta H^{d}$ = 49.1 kJ/mol | |
| | | | | | | | $\Delta S^{d}$ = 126 J/molK | |
| | | | | | | | $-\Delta H^{a}$ = 45.3 kJ/mol | |
| | | | | | | | $-\Delta S^{a}$ = 151 J/molK | |
| $TiFe_{0.30}Co_{0.70}$ | 50.0 | 15.0 | 35.0 | | | 1.10 | | [137] |
| | | | | | | RT, 30 bar | | |
| $TiFe_{0.05}Co_{0.95}$ | 50.0 | 2.5 | 47.5 | | | 1.10<br>RT, 30 bar | | [137] |
| **-Ni** | | | | | | | | |
| $Ti_{1.10}Fe_{0.90}Ni_{0.10}$ | 52.4 | 42.9 | 4.8 | $Ti_2Ni$, $TiNi_3$ | | | 40, 60, 80 °C<br>$\Delta H^d$ = 17.9 kJ/mol<br>$\Delta S^d$ = 57 J/molK<br>$-\Delta H^a$ = 16.6 kJ/mol<br>$-\Delta S^a$ = 59 J/molK | [146] |
| $TiFeNi_{0.07}$ | 48.3 | 48.3 | 3.4 | $TiFe_2$ | | | 40 °C | [119] |
| $TiFeNi_{0.50}$ | 40.0 | 40.0 | 20.0 | C14 | | | 60 °C | [145] |
| $TiFe_{0.90}Ni_{0.10}$ | 50.0 | 45.0 | 5.0 | | 2.981 | | $\Delta H^d$ = 34.8 kJ/mol<br>$\Delta S^d$ = 112 J/molK<br>$-\Delta H^a$ = 31.4 kJ/mol<br>$-\Delta S^a$ = 106 J/molK | [213] |
| | | | | | | | $-\Delta H^a$ = 31.8 kJ/mol | [142] |
| | | | | | | | 60 °C | [145] |
| | | | | | | | 20, 30, 40, 50 °C | [138] |
| | | | | $TiFe_2$, $Ti_2Fe$ | 2.977 | | 40, 60, 80, 100 °C<br>$\Delta H$ = 22.3 kJ/mol<br>$\Delta S$ = 85 J/molK | [147] |
| $TiFe_{0.85}Ni_{0.15}$ | 50.0 | 42.5 | 7.5 | | 2.984 | | $\Delta H^d$ = 37.7 kJ/mol<br>$\Delta S^d$ = 118 J/molK<br>$-\Delta H^a$ = 32.2 kJ/mol<br>$-\Delta S^a$ = 103 J/molK | [213] |
| | | | | | | | 50 °C | [138] |
| $TiFe_{0.80}Ni_{0.20}$ | 50.0 | 40.0 | 10.0 | | | 1.30 | $\Delta H_{1st}$ = 41.2 kJ/mol<br>$\Delta S_{1st}$ = 119 J/molK | [185] |
| | | | | | | | 50, 80 °C<br>$\Delta H^d$ = 41.9 kJ/mol<br>$\Delta S^d$ = 118 J/molK<br>$-\Delta H^a$ = 39.4 kJ/mol<br>$-\Delta S^a$ = 113 J/molK | [213] |
| | | | | | | | 62.5, 91, 122 °C [d]<br>$-\Delta H^a$ = 39.7 kJ/mol<br>$-\Delta S^a$ = 116 J/molK | [210] |
| | | | | | | | 55 °C [a, 30 cycles]<br>$-\Delta H^a$ = 37.2 kJ/mol | [148] |
| | | | | | | | 60 °C | [145] |
| | | | | | | 1.50<br>28 °C, 20 bar | 150 °C<br>$-\Delta H^a$ = 41.0 kJ/mol | [150] |
| | | | | | | | 50 °C | [138] |
| | | | | $TiFe_2$, $Ti_2Fe$ | 2.984 | | 40, 60, 80, 100 °C<br>$\Delta H$ = 25.1 kJ/mol<br>$\Delta S$ = 76 J/molK | [147] |
| $TiFe_{0.60}Ni_{0.40}$ | 50.0 | 30.0 | 20.0 | | | | 60 °C | [145] |
| | | | | $TiFe_2$, $Ti_2Fe$ | 2.992 | | 40, 60, 80, 100 °C<br>$\Delta H$ = 26.4 kJ/mol<br>$\Delta S$ = 58 J/molK | [147] |
| $TiFe_{0.50}Ni_{0.50}$ | 50.0 | 25.0 | 25.0 | | 3.001 | | | [237] |
| | | | | | | | $\Delta H^d$ = 45.2 kJ/mol<br>$\Delta S^d$ = 107 kJ/molK<br>$-\Delta H^a$ = 43.1 kJ/mol<br>$-\Delta S^a$ = 102 J/molK | [213] |
| $TiFe_{0.40}Ni_{0.60}$ | 50.0 | 20.0 | 30.0 | | | | 60 °C | [145] |
| $TiFe_{0.25}Ni_{0.75}$ | 50.0 | 12.5 | 37.5 | | 3.010 | | | [144,237,238] |
| $Ti_{0.90}FeNi_{0.10}$ | 45.0 | 50.0 | 5.0 | C14 | | | 60 °C | [145] |
| $Ti_{0.80}FeNi_{0.20}$ | 40.0 | 50.0 | 10.0 | C14 | | | 60 °C | [145] |
| **-Pd** | | | | | | | | |
| TiFe + <1 wt.% Pd | | | | | | | RT | [156] |
| $TiFe_{0.90}Pd_{0.05}$ | 51.3 | 46.2 | 2.7 | | 2.980 | | 0 °C [d] | [151] |
| $TiFe_{0.90}Pd_{0.10}$ | 50.0 | 45.0 | 5.0 | | 3.000 | | 0 °C [d] | [151] |
| $TiFe_{0.80}Pd_{0.20}$ | 50.0 | 40.0 | 10.0 | | 3.030 | | 0 °C [d] | [151] |
| **-Cu** | | | | | | | | |
| $TiFeCu_{0.04}$ | 49.0 | 49.0 | 2.0 | | 2.978 | 1.46<br>22°C, 40 bar | 22 °C | [117] |
| $TiFeCu_{0.10}$ | 47.6 | 47.6 | 4.8 | | | | 40 °C [d] | [136] |
| $TiFeCu_{0.11}$ | 47.5 | 47.5 | 5.0 | | 2.982 | | | [117] |
| $TiFeCu_{0.17}$ | 46.0 | 46.0 | 8.0 | | 2.982 | | | [117] |
| $TiFeCu_{0.22}$ | 45.0 | 45.0 | 10.0 | $TiFe_2$ | 2.980 | | | [117] |
| $TiFe_{0.80}Cu_{0.20}$ | 50.0 | 40.0 | 10.0 | | | | 55, 87, 116 °C [d]<br>$-\Delta H^a$ = 37.2 kJ/mol<br>$-\Delta S^a$ = 111 J/molK | [210] |
| $Ti_{0.98}FeCu_{0.02}$ + 0.5 wt.% $Fe_2O_3$ | | | | $TiFe_2$,<br>$Ti_{10}Fe_7O_3$ | | | 30 °C | [127] |
| $Ti_{0.96}FeCu_{0.04}$ + 0.5 wt.% $Fe_2O_3$ | | | | $TiFe_2$,<br>$Ti_{10}Fe_7O_3$ | | | 30 °C | [127] |
| $Ti_{0.94}FeCu_{0.06}$ + 0.5 wt.% $Fe_2O_3$ | | | | $TiFe_2$,<br>$Ti_{10}Fe_7O_3$ | | | 30 °C | [127] |
| $Ti_{0.92}FeCu_{0.08}$ + 0.5 wt.% $Fe_2O_3$ | | | | $TiFe_2$,<br>$Ti_{10}Fe_7O_3$ | | | 30 °C | [127] |
| $Ti_{0.90}FeCu_{0.10}$ + 0.5 wt.% $Fe_2O_3$ | | | | $TiFe_2$, | | | 30 °C | [127] |
| | | | | $Ti_{10}Fe_7O_3$ | | | | |
| **-Y** | | | | | | | | |
| $TiFeY_{0.04}$ | 48.9 | 48.9 | 2.2 | $TiFe_2$, Y | | 1.79<br>RT, 25 bar | | [157] |
| $TiFeY_{0.07}$ | 48.3 | 48.3 | 3.4 | $TiFe_2$, Y | | 1.75<br>RT, 25 bar | | [157] |
| $TiFeY_{0.09}$ | 47.8 | 47.8 | 4.4 | β-Ti, Y | | 1.71<br>RT, 25 bar | | [157] |
| **-La** | | | | | | | | |
| $TiFe_{0.78}La_{0.03}$ | 55.4 | 43.2 | 1.4 | Ti, La | | | 20, 30, 65 °C<br>$\Delta H^d = 26.7$ kJ/mol<br>$-\Delta H^a = 24.8$ kJ/mol | [158] |
| **-Ce** | | | | | | | | |
| $TiFe_{0.94}Ce_{0.06}$ | 50.0 | 47.0 | 3.0 | | 2.983 | | | [159] |
| **-Mm** | | | | | | | | |
| TiFe + $_{4\ wt.\%}$ Mm | | | | | | | 27 °C $^d$ | [160] |
| $Ti_{1.3}Fe$ + $_{1.5\ wt.\%}$ Mm | | | | β-Ti | | | 25, 45, 60 °C $^d$ | [161] |
| $Ti_{1.3}Fe$ + $_{4.0\ wt.\%}$ Mm | | | | β-Ti | | | 25, 45, 60 °C $^d$ | [161] |
| $Ti_{1.3}Fe$ + $_{4.5\ wt.\%}$ Mm | | | | Ti, $Ti_2Fe$ | 2.978 | 1.71, RT<br>1.90, 100°C<br>20 bar | RT, 100 °C $^d$ | [162] |
| $Ti_{1.4}Fe$ + $_{4.5\ wt.\%}$ Mm | | | | Ti, $Ti_2Fe$ | 2.978 | 1.85<br>100°C, 20 bar | 100 °C $^d$ | [162] |
| $Ti_{1.5}Fe$ + $_{4.5\ wt.\%}$ Mm | | | | Ti, $Ti_2Fe$ | 2.978 | 1.17<br>100°C, 20 bar | 100 °C $^d$ | [162] |
| **-Al** | | | | | | | | |
| $TiFeAl_{0.11}$ | 47.5 | 47.5 | 5.0 | | 2.980 | 0.99<br>22 °C, 60 bar | 22 °C | [134] |
| $TiFeAl_{0.22}$ | 45.0 | 45.0 | 10.0 | | 2.982 | | | [134] |
| $TiFeAl_{0.35}$ | 42.5 | 42.5 | 15.0 | | 2.988 | | | [134] |
| $TiFeAl_{0.50}$ | 40.0 | 40.0 | 20.0 | | 2.991 | | | [134] |
| $TiFe_{0.98}Al_{0.02}$ | 50.0 | 49.0 | 1.0 | | | | 30 °C<br>$\Delta H_{1st}{}^d = 26.4$ kJ/mol<br>$\Delta S_{1st}{}^d = 100$ J/molK<br>$-\Delta H_{1st}{}^a = 23.8$ kJ/mol<br>$-\Delta S_{1st}{}^a = 98$ J/molK | [239] |
| | | | | | 2.977 | | 50 °C $^d$ | [138] |
| $TiFe_{0.96}Al_{0.04}$ | 50.0 | 48.0 | 2.0 | | 2.988 | | 40 °C $^d$ | [122] |
| | | | | | | | 30 °C<br>$\Delta H_{1st}{}^d = 23.4$ kJ/mol<br>$\Delta S_{1st}{}^d = 95$ J/molK | [239] |
| | | | | | | | -$\Delta H_{1st}$ [a] = 21.9 kJ/mol | |
| | | | | | | | -$\Delta S_{1st}$ [a] = 97 J/molK | |
| | | | | | 2.986 | 1.10 | 25,[d] 50, 80 °C | [110] |
| | | | | | | 50 °C, 10 bar | $\Delta H$ = 29.9 kJ/mol | |
| | | | | | | | $\Delta S$ = 106 J/molK | |
| $TiFe_{0.95}Al_{0.05}$ | 50.0 | 47.5 | 2.5 | | 2.985 | | 50 °C [d] | [138] |
| $TiFe_{0.94}Al_{0.06}$ | 50.0 | 47.0 | 3.0 | | | | 30, 40, 50 °C | [164] |
| | | | | | | | 30 °C | [239] |
| | | | | | | | $\Delta H_{1st}$ [d] = 21.4 kJ/mol | |
| | | | | | | | $\Delta S_{1st}$ [d] = 89 J/molK | |
| | | | | | | | -$\Delta H_{1st}$ [a] = 19.6 kJ/mol | |
| | | | | | | | -$\Delta S_{1st}$ [a] = 88 J/molK | |
| $TiFe_{0.92}Al_{0.08}$ | 50.0 | 46.0 | 4.0 | | | | 30 °C | [239] |
| | | | | | | | $\Delta H_{1st}$ [d] = 29.5 kJ/mol | |
| | | | | | | | $\Delta S_{1st}$ [d] = 109 J/molK | |
| | | | | | | | -$\Delta H_{1st}$ [a] = 22.7 kJ/mol | |
| | | | | | | | -$\Delta S_{1st}$ [a] = 91 J/molK | |
| $TiFe_{0.90}Al_{0.10}$ | 50.0 | 45.0 | 5.0 | | 2.989 | | 40 °C [d] | [122] |
| | | | | | | | 30 °C | [239] |
| | | | | | | | $\Delta H_{1st}$ [d] = 20.8 kJ/mol | |
| | | | | | | | $\Delta S_{1st}$ [d] = 78 J/molK | |
| | | | | | | | -$\Delta H_{1st}$ [a] = 18.3 kJ/mol | |
| | | | | | | | -$\Delta S_{1st}$ [a] = 73 J/molK | |
| | | | | | 2.997 | | 50 °C [d] | [138] |
| | | | | | 2.996 | 1.00 | 25,[d] 50, 80 °C | [110] |
| | | | | | | 50 °C, 10 bar | $\Delta H$ = 31.2 kJ/mol | |
| | | | | | | | $\Delta S$ = 105 J/molK | |
| $TiFe_{0.80}Al_{0.20}$ | 50.0 | 40.0 | 10.0 | | 3.015 | 0.90 | 50, 80, 120[d] °C | [110] |
| | | | | | | 50 °C, 10 bar | $\Delta H$ = 35.0 kJ/mol | |
| | | | | | | | $\Delta S$ = 100 J/molK | |
| $TiFe_{0.76}Al_{0.24}$ | 50.0 | 38.0 | 12.0 | | 3.030 | | 40 °C [d] | [122] |
| $TiFe_{0.60}Al_{0.40}$ | 50.0 | 30.0 | 20.0 | | 3.070 | | | [122] |
| **-Si** | | | | | | | | |
| $TiFe_{0.94}Si_{0.02}$ | 51.0 | 48.0 | 1.0 | $TiFe_2$, β-Ti | | | 40 °C [d] | [122] |
| $TiFe_{0.87}Si_{0.03}$ | 52.0 | 45.0 | 3.0 | $TiFe_2$, β-Ti | | | 40 °C [d] | [122] |
| $TiFe_{0.83}Si_{0.10}$ | 52.0 | 43.0 | 5.0 | $TiFe_2$, β-Ti | | | 40 °C [d] | [122] |
| **-Sn** | | | | | | | | |
| $TiFe_{0.98}Sn_{0.02}$ | 50.0 | 49.0 | 1.0 | $TiFe_2$ | | 1.60 | RT,[a,d] 30, 40, 50, 60 °C [d] | [165] |
| | | | | | | | $\Delta H$ = 27.0 kJ/mol | |
| | | | | | | | $\Delta S$ = 103 J/molK | |
| $TiFe_{0.95}Sn_{0.05}$ | 50.0 | 47.5 | 2.5 | $TiFe_2$ | | 1.40 | RT,[a,d] 30, 40, 50, 60 °C [d] | [165] |
| | | | | | | | ΔH = 26.6 kJ/mol<br>ΔS = 103 J/molK | |
| **-B** | | | | | | | | |
| $TiFeB_{0.001}$ | 49.98 | 49.97 | 0.05 | $TiFe_2$, Ti | | | 50 °C | [166] |
| $Ti_{1.1}FeB_{0.001}$ | 52.35 | 47.60 | 0.05 | $TiFe_2$, Ti | | | 50 °C | [166] |
| **-C** | | | | | | | | |
| $TiFeC_{0.001}$ | 49.98 | 49.97 | 0.05 | $TiFe_2$, Ti | | | 50 °C | [166] |
| $Ti_{1.1}FeC_{0.001}$ | 52.35 | 47.60 | 0.05 | $TiFe_2$, Ti | | | 50 °C | [166] |
| **-S** | | | | | | | | |
| $TiFe_{0.95}S_{0.02}$ | 50.7 | 48.3 | 1.0 | $Ti_2S$ | | | 25, 50 °C<br>-ΔH[a] = 21.8 kJ/mol | [176] |
| $TiFeS_{0.02}$ | 49.5 | 49.5 | 1.0 | $Ti_2S$ | 2.978 | 0.95<br>22 °C, 40 bar | 22 °C | [108] |

**Table 3 –** Alloy composition, secondary phases, TiFe-phase lattice parameter and main hydrogenation properties of monosubstituted Ti(Fe,Mn) systems . [a]: absorption, [d]: desorption, $_{1st}$: 1$^{st}$ plateau, $_{2nd}$: 2$^{nd}$ plateau, RT: room temperature.

| Stoichiometry | Ti %at | Fe %at | Mn %at | Secondary Phases | TiFe a (Å) | Capacity (%wt) Conditions | PCI Thermodynamics | Ref. |
|---|---|---|---|---|---|---|---|---|
| **Ti:Fe 1:1** | | | | | | | | |
| $TiFe_{0.95}Mn_{0.05}$ | 50.0 | 47.5 | 2.5 | | | | 5, 50 °C $^d$<br>$\Delta H^d$ = 29.3 kJ/mol<br>$\Delta S^d$ = 108 J/molK<br>$-\Delta H^a$ = 25.1 kJ/mol<br>$-\Delta S^a$ = 101 J/molK | [213] |
| | | | | | 2.977 | 1.73<br>25°C, 80 bar | 25, 55, 85 °C<br>$\Delta H_{1st}{}^d$ = 28.2 kJ/mol<br>$\Delta S_{1st}{}^d$ = 105 J/molK<br>$-\Delta H_{1st}{}^a$ = 24.7 kJ/mol<br>$-\Delta S_{1st}{}^a$ = 100 J/molK | [178] |
| $TiFe_{0.90}Mn_{0.10}$ | 50.0 | 45.0 | 5.0 | | | 1.68<br>25°C, 40 bar | 8, 22, 45 °C $^d$<br>$-\Delta H^a$ = 21.8 kJ/mol<br>$-\Delta S^a$ = 89 J/molK<br>$\Delta H^d$ = 30.3 kJ/mol<br>$\Delta S^d$ = 110 J/molK | [187] |
| | | | | | 2.980 | | $\Delta H^d$ = 27.6 kJ/mol<br>$\Delta S^d$ = 100 J/molK<br>$-\Delta H^a$ = 23.9 kJ/mol<br>$-\Delta S^a$ = 95 J/molK | [213] |
| | | | | | | | 50 °C $^d$ | [131] |
| | | | | | | | 40 °C $^d$ | [136] |
| | | | | | | | 25 °C<br>$-\Delta H^a$ = 23.9 kJ/mol | [52] |
| | | | | | | | 30, 40, 50 °C | [240] |
| | | | | | | | 25, 40 °C | [201] |
| | | | | | | 1.70 | | [183] |
| | | | | | | | 40 °C | [241] |
| | | | | Ti-type | 2.979 | | 10, 30, 50 °C<br>$\Delta H^d$ = 29.3 kJ/mol<br>$\Delta S^d$ = 105 kJ/molK<br>$-\Delta H^a$ = 24.2 kJ/mol<br>$-\Delta S^a$ = 96 J/molK | [132] |
| | | | | | 2.979 | 1.75<br>25°C, 79 bar | 25, 55, 85 °C<br>$\Delta H_{1st}{}^d$ = 28.1 kJ/mol | [178] |
| | | | | | | | $\Delta S_{1st}$ [d] = 102 J/molK | |
| | | | | | | | $-\Delta H_{1st}$ [a] = 24.7 kJ/mol | |
| | | | | | | | $-\Delta S_{1st}$ [a] = 97 J/molK | |
| | | | | | | | $\Delta H_{2nd}$ [d] = 38.9 kJ/mol | |
| | | | | | | | $\Delta S_{2nd}$ [d] = 144 J/molK | |
| | | | | | | | $-\Delta H_{2nd}$ [a] = 29.7 kJ/mol | |
| | | | | | | | $-\Delta S_{2nd}$ [a] = 120 J/molK | |
| $TiFe_{0.885}Mn_{0.115}$ | 50.0 | 44.2 | 5.8 | | 2.985 | | | [70] |
| $TiFe_{0.85}Mn_{0.15}$ | 50.0 | 42.5 | 7.5 | | | 1.90 | $\Delta H_{1st}$ = 29.5 kJ/mol | [185] |
| | | | | | | | $\Delta S_{1st}$ = 107 J/molK | |
| | | | | | | 1.70 | 40, 50 [d], 60 [d], 70 [d] °C | [136,183] |
| | | | | | | | 30 °C | [48] |
| | | | | | | | $-\Delta H^{a}$ = 32.6 kJ/mol | [142] |
| | | | | | | | 20, 30, 40, 50 °C [d] | [186] |
| $TiFe_{0.80}Mn_{0.20}$ | 50.0 | 40.0 | 10.0 | | | | 5 °C | [213] |
| | | | | | | | $\Delta H$ [d] = 31.8 kJ/mol | |
| | | | | | | | $\Delta S$ [d] = 105 J/molK | |
| | | | | | | | $-\Delta H$ [a] = 27.6 kJ/mol | |
| | | | | | | | $-\Delta S$ [a] = 95 J/molK | |
| | | | | | | | 18, 45, 80 °C | [210] |
| | | | | | | | $-\Delta H$ = 21.8 kJ/mol | |
| | | | | | | | $-\Delta S$ = 75 J/molK | |
| | | | | $TiMn_{1.5}$ | | | 50 °C [d] | [131] |
| | | | | | 2.983 | | 22 °C [d] | [195] |
| | | | | | | | 61 °C [d] | [181] |
| | | | | | | | 40 °C [d] | [136] |
| | | | | | | 1.65 | 25 °C | [52] |
| | | | | | | 25°c, 41 bar | $-\Delta H$ [a] = 28.4 kJ/mol | |
| $TiFe_{0.70}Mn_{0.30}$ | 50.0 | 35.0 | 15.0 | | 2.994 | | 22 °C | [195] |
| | | | | | | | 40 °C [d] | [136,181] |
| | | | | | | | 30 °C | [48] |
| | | | | | | | 25 °C | [52] |
| | | | | | | | $-\Delta H$ [a] = 29.3 kJ/mol | |
| **Ti-rich side** | | | | | | | | |
| $TiFe_{0.88}Mn_{0.02}$ | 52.6 | 46.3 | 1.1 | β-Ti, $Ti_4Fe_2O$ | 2.986 | 1.86 | 25, 40, 55 °C | [200] |
| | | | | | | 25 °C, 55 bar | $-\Delta H_{1st}$ [a] = 29.8 kJ/mol | |
| | | | | | | | $-\Delta S_{1st}$ [a] = 106 J/molK | |
| | | | | | | | $\Delta H_{1st}$ [d] = 33.3 kJ/mol | |
| | | | | | | | $\Delta S_{1st}$ [d] = 112 J/molK | |
| | | | | | | | $-\Delta H_{2nd}$ [a] = 30.3 kJ/mol | |
| | | | | | | | $-\Delta S_{2nd}$ [a] = 112 J/molK | |
| | | | | | | | $\Delta H_{2nd}^{d}$ = 27.2 kJ/mol<br>$\Delta S_{2nd}^{d}$ = 106 J/molK | |
| $TiFe_{0.85}Mn_{0.05}$ | 52.6 | 44.7 | 2.6 | β-Ti, $Ti_4Fe_2O$ | 2.985 | 1.73<br>25°C, 24 bar | 25, 40, 55, 70 °C<br>$\Delta H_{1st}^{d}$ = 30.6 kJ/mol<br>$\Delta S_{1st}^{d}$ = 103 J/molK<br>$-\Delta H_{1st}^{a}$ = 27.8 kJ/mol<br>$-\Delta S_{1st}^{a}$ = 99 J/molK<br>$\Delta H_{2nd}^{d}$ = 35.2 kJ/mol<br>$\Delta S_{2nd}^{d}$ = 126 J/molK<br>$-\Delta H_{2nd}^{a}$ = 32.5 kJ/mol<br>$-\Delta S_{2nd}^{a}$ = 121 J/molK | [178] |
| $TiFe_{0.80}Mn_{0.10}$ | 52.6 | 42.1 | 5.3 | $TiFe_2$, β-Ti | 2.988 | 1.92<br>25°C, 20 bar | 25, 65 °C | [141] |
| | | | | $TiFe_2$, β-Ti | 2.989 | 1.68<br>25°C, 25 bar | 25°C | [236] |
| | | | | β-Ti, $Ti_4Fe_2O$ | 2.987 | 1.77<br>25 °C, 55 bar | 25, 40, 55 °C<br>$\Delta H_{1st}^{d}$ = 29.2 kJ/mol<br>$\Delta S_{1st}^{d}$ = 96 J/molK<br>$-\Delta H_{1st}^{a}$ = 28.3 kJ/mol<br>$-\Delta S_{1st}^{a}$ = 98 J/molK<br>$\Delta H_{2nd}^{d}$ = 35.4 kJ/mol<br>$\Delta S_{2nd}^{d}$ = 124 J/molK<br>$-\Delta H_{2nd}^{a}$ = 31.1 kJ/mol<br>$-\Delta S_{2nd}^{a}$ = 114 J/molK | [178] |
| $TiFe_{0.70}Mn_{0.20}$ | 52.6 | 36.9 | 10.5 | $TiFe_2$, β-Ti | 2.993 | 1.98<br>25°C, 20 bar | 25, 45,[d] 65 [d] °C | [141] |
| | | | | | | | 30 °C<br>$-\Delta H_{1st}^{a}$ = 36.3 kJ/mol<br>$-\Delta S_{1st}^{a}$ = 106 J/molK<br>$-\Delta H_{2nd}^{a}$ = 32.1 kJ/mol<br>$-\Delta S_{2nd}^{a}$ = 111 J/molK | [105,136,189] |
| $TiFe_{0.90}Mn_{0.05}$ | 51.3 | 46.2 | 2.6 | $Ti_4Fe_2O$ | 2.982 | 1.84<br>25°C, 55 bar | 25, 40, 55 °C<br>$\Delta H_{1st}^{d}$ = 28.2 kJ/mol<br>$\Delta S_{1st}^{d}$ = 98 J/molK<br>$-\Delta H_{1st}^{a}$ = 26.2 kJ/mol<br>$-\Delta S_{1st}^{a}$ = 97 J/molK<br>$\Delta H_{2nd}^{d}$ = 35.3 kJ/mol<br>$\Delta S_{2nd}^{d}$ = 128 J/molK<br>$-\Delta H_{2nd}^{a}$ = 31.1 kJ/mol<br>$-\Delta S_{2nd}^{a}$ = 120 J/molK | [178] |
| $TiFe_{0.80}Mn_{0.05}$ | 54.1 | 43.2 | 2.7 | β-Ti, $Ti_4Fe_2O$ | 2.986 | 1.55 | 25, 40, 55 °C | [178] |
| | | | | | | 25°C, 57 bar | $\Delta H_{1st}^{d}$ = 32.4 kJ/mol<br>$\Delta S_{1st}^{d}$ = 107 J/molK<br>$-\Delta H_{1st}^{a}$ = 27.6 kJ/mol<br>$-\Delta S_{1st}^{a}$ = 98 J/molK<br>$\Delta H_{2nd}^{d}$ = 38.9 kJ/mol<br>$\Delta S_{2nd}^{d}$ = 138 J/molK<br>$-\Delta H_{2nd}^{a}$ = 35.4 kJ/mol<br>$-\Delta S_{2nd}^{a}$ = 132 J/molK | |
| $TiFe_{0.86}Mn_{0.10}$ | 51.0 | 43.9 | 5.1 | | 2.984 | 1.94<br>0 °C, 50 bar | 45, 60, 80 °C $^{d}$<br>$\Delta H^{d}$ = 29.5 kJ/mol<br>$\Delta S^{d}$ = 106 J/molK | [198] |
| | | | | | 2.978 | 1.83 | 60 °C | [242] |
| $TiFe_{0.76}Mn_{0.13}$ | 52.9 | 40.2 | 6.9 | | | | 40 °C $^{d}$ | [136] |
| $Ti_{1.10}Fe_{0.80}Mn_{0.20}$ | 52.4 | 38.1 | 9.5 | | | | 40, 60, 80 °C<br>$\Delta H^{d}$ = 23.6 kJ/mol | [243] |
| **Fe-rich side** | | | | | | | | |
| $TiFeMn_{0.04}$ | 49.0 | 49.0 | 2.0 | | 2.970 | | | [117] |
| $TiFeMn_{0.05}$ | 48.8 | 48.8 | 2.4 | $TiFe_2$ | 2.976 | 1.55<br>25°C, 58 bar | 25, 55, 85 °C<br>$\Delta H_{1st}^{d}$ = 28.2 kJ/mol<br>$\Delta S_{1st}^{d}$ = 105 J/molK<br>$-\Delta H_{1st}^{a}$ = 24.3 kJ/mol<br>$-\Delta S_{1st}^{a}$ = 99 J/molK<br>$\Delta H_{2nd}^{d}$ = 35.3 kJ/mol<br>$\Delta S_{2nd}^{d}$ = 139 J/molK<br>$-\Delta H_{2nd}^{a}$ = 26.2 kJ/mol<br>$-\Delta S_{2nd}^{a}$ = 116 J/molK | [178] |
| $TiFeMn_{0.08}$ | 48.0 | 48.0 | 4.0 | | 2.976 | | | [117] |
| $TiFeMn_{0.13}$ | 47.0 | 47.0 | 6.0 | | 2.977 | 1.14<br>22°C, 32 bar | 22 °C | [117] |
| $TiFeMn_{0.17}$ | 46.0 | 46.0 | 8.0 | | 2.977 | | | [117] |
| $TiFe_{0.94}Mn_{0.10}$ | 49.0 | 46.0 | 5.0 | | 2.976 | | 22 °C $^{d}$ | [195] |
| $TiFe_{0.83}Mn_{0.25}$ | 48.0 | 40.0 | 12.0 | | 2.990 | | 22 °C $^{d}$ | [195] |

**Table 4** – Details on the investigated composition in the Ti(Fe,Mn)-M system. [a]: absorption, [d]: desorption, $_{1st}$: 1$^{st}$ plateau, $_{2nd}$: 2$^{nd}$ plateau, RT: room temperature.

| TiFeMn-M | Ti %at | Fe %at | Mn %at | M %at | Secondary Phases | TiFe a (Å) | Capacity (%wt) Conditions | PCI Thermodynamics | Ref. |
|---|---|---|---|---|---|---|---|---|---|
| **-Zr** | | | | | | | | | |
| $TiFeMn_{0.02}Zr_{0.01}$ | 49.2 | 49.2 | 1.0 | 0.6 | ($TiFe_2$) | 2.980 | | | [194] |
| $TiFeMn_{0.04}Zr_{0.02}$ | 48.5 | 48.5 | 2.0 | 1.0 | | 2.980 | 1.00<br>22 °C, 40 bar | | [191] |
| | | | | | ($TiFe_2$) | 2.980 | | RT | [194] |
| $TiFeMn_{0.09}Zr_{0.04}$ | 47.0 | 47.0 | 4.0 | 2.0 | ($TiFe_2$) | 2.980 | | RT | [194] |
| $TiFeMn_{0.12}Zr_{0.06}$ | 45.8 | 45.8 | 5.6 | 2.8 | ($TiFe_2$) | 2.986 | | RT | [194] |
| **-V** | | | | | | | | | |
| Ti-Fe-Mn-V | 45.7 | 44.7 | 8.4 | 1.2 | C14 | 2.971 | | 22 °C $^d$ | [195] |
| Ti-Fe-Mn-V | 45.7 | 44.7 | 7.2 | 2.4 | C14 | 2.978 | | 22 °C $^d$ | [195] |
| Ti-Fe-Mn-V | 45.7 | 44.7 | 4.8 | 4.8 | C14 | 2.981 | | 22, 30, 50 ,80 °C $^d$<br>$\Delta H_{1st}{}^d$ = 28.6 kJ/mol<br>$\Delta S_{1st}{}^d$ = 104 J/molK<br>$\Delta H_{2nd}{}^d$ = 44.9 kJ/mol<br>$\Delta S_{2nd}{}^d$ = 159 J/molK | [195,196] |
| Ti-Fe-Mn-V | 45.7 | 44.7 | 1.2 | 8.4 | C14 | 2.983 | | 22 °C $^d$ | [195] |
| Ti-Fe-Mn-V | 41.7 | 41.0 | 13.0 | 4.3 | C14 | 2.987 | | 22 °C $^d$ | [195] |
| Ti-Fe-Mn-V | 48.0 | 37.0 | 10.0 | 5.0 | C14 | 2.988 | | 22 °C $^d$ | [195] |
| Ti-Fe-Mn-V | 43.6 | 42.8 | 9.1 | 4.5 | C14 | 2.990 | | 22 °C $^d$ | [195] |
| Ti-Fe-Mn-V | 40.0 | 39.1 | 16.7 | 4.2 | C14 | 2.994 | | 22 °C $^d$ | [195] |
| $TiFe_{0.91}Mn_{0.10}V_{0.01}$ | 49.5 | 45.0 | 5.0 | 0.5 | | | | 25 °C $^d$ | [235] |
| $TiFe_{0.82}Mn_{0.20}V_{0.02}$ | 49.0 | 40.0 | 10.0 | 1.0 | | | | 25 °C $^d$ | [235] |
| $TiFe_{0.92}Mn_{0.10}V_{0.02}$ | 49.0 | 45.0 | 5.0 | 1.0 | | | | 25 °C $^d$ | [235] |
| $TiFe_{0.80}Mn_{0.10}V_{0.05}$ | 51.3 | 41.0 | 5.1 | 2.6 | $TiFe_2$, β-Ti | 2.994 | 1.71<br>25°C, 25 bar | 25 °C | [236] |
| $TiFe_{0.80}Mn_{0.10}V_{0.10}$ | 50.0 | 40.0 | 5.0 | 5.0 | $TiFe_2$, β-Ti | 2.996 | 1.76<br>25°C, 25 bar | 25 °C | [236] |
| $TiFe_{0.80}Mn_{0.20}$ + V | | | | | | | | 50, 60, 70 °C<br>ΔH = 33.4 kJ/mol<br>ΔH = 31.8 kJ/mol | [197] |
| **-Co** | | | | | | | | | |
| $TiFe_{0.86}Mn_{0.06}Co_{0.04}$ | 51.0 | 43.9 | 3.1 | 2.0 | | 2.983 | 1.97<br>0 °C, 50 bar | 45, 60, 80 °C $^d$<br>$\Delta H^d$ = 28.5 kJ/mol<br>$\Delta S^d$ = 102 J/molK | [198] |

| | | | | | | | | |
|---|---|---|---|---|---|---|---|---|
| $TiFe_{0.86}Mn_{0.05}Co_{0.05}$ | 51.0 | 43.9 | 2.6 | 2.6 | | 2.983 | 1.98<br>0 °C, 50 bar | 45, 60, 80 °C [d]<br>$\Delta H^{d}$ = 29.3 kJ/mol<br>$\Delta S^{d}$ = 104 J/molK | [198] |
| $TiFe_{0.86}Mn_{0.04}Co_{0.06}$ | 51.0 | 43.9 | 2.0 | 3.1 | | 2.983 | 1.96<br>0 °C, 50 bar | 45, 60, 80 °C [d]<br>$\Delta H^{d}$ = 29.8 kJ/mol<br>$\Delta S^{d}$ = 106 J/molK | [198] |
| $TiFe_{0.86}Mn_{0.05}Co_{0.08}$ | 50.3 | 43.2 | 2.5 | 4.0 | | 2.982 | | 45 °C [d] | [198] |
| $TiFe_{0.86}Mn_{0.04}Co_{0.10}$ | 50.0 | 43.0 | 2.0 | 5.0 | | 2.980 | | 30, 45 [d], 50, 55 °C | [198] |
| $TiFe_{0.86}Mn_{0.05}Co_{0.12}$ | 49.3 | 42.4 | 2.5 | 5.9 | | 2.981 | | 45 °C [d] | [198] |
| **-Ni** | | | | | | | | | |
| $TiFe_{0.70}Mn_{0.16}Ni_{0.04}$ | 52.6 | 36.9 | 8.4 | 2.1 | | 2.993 | 1.80<br>25°C, 20 bar | | [141] |
| $TiFe_{0.70}Mn_{0.12}Ni_{0.08}$ | 52.6 | 36.9 | 6.3 | 4.2 | | 2.994 | 1.66<br>25°C, 20 bar | 25 °C [a] | [141] |
| **-Cu** | | | | | | | | | |
| $TiFe_{0.86}Mn_{0.02}Cu_{0.02}$ | 52.6 | 45.2 | 1.1 | 1.1 | β-Ti,<br>$Ti_4Fe_2O$ | 2.988 | 1.83<br>25 °C, 55 bar | 25, 40, 55 °C<br>$-\Delta H_{1st}^{a}$ = 34.1 kJ/mol<br>$-\Delta S_{1st}^{a}$ = 115 J/molK<br>$\Delta H_{1st}^{d}$ = 37.6 kJ/mol<br>$\Delta S_{1st}^{d}$ = 122 J/molK<br>$-\Delta H_{2nd}^{a}$ = 29.3 kJ/mol<br>$-\Delta S_{2nd}^{a}$ = 115 J/molK<br>$\Delta H_{2nd}^{d}$ = 32.5 kJ/mol<br>$\Delta S_{2nd}^{d}$ = 122 J/molK | [200] |
| $TiFe_{0.84}Mn_{0.02}Cu_{0.04}$ | 52.6 | 44.2 | 1.1 | 2.1 | β-Ti,<br>$Ti_4Fe_2O$ | 2.992 | 1.68<br>25 °C, 55 bar | 25, 40, 55 °C<br>$-\Delta H_{1st}^{a}$ = 33.4 kJ/mol<br>$-\Delta S_{1st}^{a}$ = 107 J/molK<br>$\Delta H_{1st}^{d}$ = 34.7 kJ/mol<br>$\Delta S_{1st}^{d}$ = 107 J/molK<br>$-\Delta H_{2nd}^{a}$ = 25.1 kJ/mol<br>$-\Delta S_{2nd}^{a}$ = 107 J/molK<br>$\Delta H_{2nd}^{d}$ = 26.9 kJ/mol<br>$\Delta S_{2nd}^{d}$ = 107 J/molK | [200] |
| **-Y** | | | | | | | | | |
| $TiFe_{0.90}Mn_{0.10}Y_{0.05}$ | 48.8 | 43.9 | 4.9 | 2.4 | α-Y | 2.981 | | 10, 30, 50 °C<br>$\Delta H^{d}$ = 31.2 kJ/mol<br>$\Delta S^{d}$ = 109 J/molK<br>$-\Delta H^{a}$ = 26.7 kJ/mol<br>$-\Delta S^{a}$ = 102 J/molK | [132] |
| **-Ce** | | | | | | | | | |
| $TiFe_{0.90}Mn_{0.10}Ce_{0.02}$ | 49.5 | 44.6 | 5.0 | 1.0 | Ce, $CeO_2$ | | | 25 °C | [201] |

| | | | | | | | | |
|---|---|---|---|---|---|---|---|---|
| $TiFe_{0.90}Mn_{0.10}Ce_{0.04}$ | 49.0 | 44.1 | 4.9 | 2.0 | Ce, $CeO_2$ | | 25, 40 °C | [201] |
| $TiFe_{0.90}Mn_{0.10}Ce_{0.06}$ | 48.5 | 43.7 | 4.9 | 2.9 | Ce, $CeO_2$ | | 25 °C | [201] |
| **-Mm** | | | | | | | | |
| $Ti_{1+x}Fe_{1-y}Mn_yMm_z$ (x=0.0-0.9, y=0.04-0.2, z=0.002-0.028) | | | | | β-Ti | 2.15 | 20, 49, 77, 99 °C [d] | [202] |

**Table 5** – Overview of effects of elemental substitution on TiFe intermetallic compounds properties. Sub.: substituent; Act.: activation; Kin.: kinetics; a: lattice constant of TiFe; β $P_{eq}$ ($1^{st}$): equilibrium pressure of the $1^{st}$ plateau related to the beta monohydride; γ $P_{eq}$ ($2^{nd}$): equilibrium pressure of the $2^{nd}$ plateau related to the gamma dihydride; Plat.: plateau; Cap.: capacity; Hys.: hysteresis; ΔH: enthalpy of hydrogenation; ΔS: entropy of hydrogenation; Cyc.: cycling; Res. to Pois.: resistance to poisoning; ✗: no/not improved/suppressed; ✓: yes/improved; ↑: higher; ↓: lower; =: not changed; II ph.: secondary phases.

| **Element(s)** | **Ti Sub.** | **Fe Sub.** | **Act.** | **Kin.** | **TiFe a** | **β $P_{eq}$ ($1^{st}$)** | **γ $P_{eq}$ ($2^{nd}$)** | **Plat.** | **Cap.** | **Hys.** | **ΔH** | **ΔS** | **Cyc.** | **Res. to Pois.** |
|---|---|---|---|---|---|---|---|---|---|---|---|---|---|---|
| **Mg** | | | ✓ | | ↑ | ↓ | ✗ | | ↓ | | | | | |
| **Be** | | ✓ | | ✓ | ↓ | ↓ | ✗ | | ↓ | ↓ | ↑ | = | | |
| **Zr** | ✓ | | ✓ | ✓ | ↑ | ↓ | ✗ | Sloped | ↓ | = | = | = | | ✓ Air |
| **Hf** | | | ✓ II ph. | ✓ | ↑ | ↓ | | | ↓ | | | | | |
| **V** | ✓ | ✓ | ✓ II ph. | ✓ | ↑ | ↓ | ↓ | Smoothed | ↑ | ↓ | | | ✗ | ✗ |
| **Nb** | ✓ | | ✓ II ph., brittle | | ↑ | ↓ | ↓ | | | | | | | ✓ |
| **Ta** | | | | | ↑ | ↓ | | | | | | | | |
| **Cr** | | ✓ | ✓ II ph., brittle | ✓ | ↑ | ↓ | ✗ | Sloped | ↓ | ↓ | ↑ | | | |
| **CrY** | ✓ | ✓ | ✓ II ph., brittle | ✓ | ↑ | ↓ | ↓ | Sloped | | ↓ | ↑ | | | |
| **CrZr** | | | ✓ II ph. | | | | | | | | | | | |
| **Mo** | | ✓ | | | | ↓ | | Sloped | | | | | | |
| **Co** | | ✓ | ✓ II ph. | | ↑ | ↓ | ↑ | | ↓ | ↓ | | | | ✓ |
| **Ni** | | ✓ | ✓ II ph., reactive surface | ✓ | ↑ | ↓ | ↑ | $2^{nd}$ Sloped | ↓ | ↓ | | | ✓ | |
| **NiZr** | ✓ | ✓ | | | ↑ | ↓ | | | ↓ | | | | | |
| **NiV** | ✓ | ✓ | ✓ | ✓ | ↑ | ↓ | | | ↑ | ↓ | = | = | | |
| **NiNb** | ✓ | ✓ | ✓ | ✓ | ↑ | ↓ | | | ↑ | ↓ | = | = | | |
| **Pd** | | | ✓ | ✓ | ↑ | ↓ | = | | | | | | | ✓ Air |
| **Cu** | | ✓ | ✓ II ph., brittle | ✓ | ↑ | ↓ | ↑ | | ↓ | | | | | |
| **Y** | ✓ | | ✓ II ph. | ✓ | ↑ | ↓ | | | ↓ | | | | | |
| **La** | ✗ | ✗ | ✓ cracks | | = | | | | ↑ | ↓ | | | | |
| **Ce** | | ✓ | ✓ reactive surface | ✓ | ↑ | | | | | | | | | |
| **Mm** | ✗ | ✗ | ✓ II ph., cracks | ✓ | = | | | | | | | | | ✓ $O_2$, $CO_2$ |
| **Al** | | ✓ | | ✓ | ↑ | ↑ | ✗ | Sloped | ↓ | ↓ | ↑ | = | | |
| **Si** | | ✓ | ✓ II ph., | | ↑ | ↓ | ↑ | Sloped | ↓ | | | | | |
| **Sn** | ✓ | ✓ | ✓ II ph. | ✓ | ↓ | ↑ | | | ↓ | | ↑ | ↑ | | |
| **B** | | | ✓ II ph. | | | ↑ | ✗ | Sloped | ↓ | | | | | |
| **C** | | | ✓ II ph. | | | ↑ | ✗ | Sloped | ↓ | | | | | |
| **N** | | | ✓ II ph. | | | | | | ↓ | | | | | |
| **S** | | | ✓ II ph. | | ↑ | ↓ | ✗ | | ↓ | | | | ✓ | ✓ |
| **Mn** | | ✓ | ✓ II ph., reactive surface | ✓ | ↑ | ↓ | ↓ | Smoothed | ↑ | ↓ | ↑ | ↑ | ✓ | ✓ CO |
| **MnZr** | ✓ | ✓ | ✓ II ph. | ✓ | ↑ | ↓ | ✗ | | ↓ | | | | | ✓ Air |
| **MnV** | ✓ | ✓ | ✓ II ph. | ✗ | ↑ | ↓ | | Smoothed | ↑ | ↓ | ↑ | | | ✗ |
| **MnCo** | | ✓ | ✓ | | ↑ | ↓ | ↑ | Sloped | ↓ | | | | ✓ | |
| **MnNi** | | ✓ | ✓ | | = | ↓ | ↑ | | ↓ | | | | ✓ | |
| **MnCu** | | ✓ | ✓ II ph. | ✓ | ↑ | ↓ | ↑ | | | | | | | |
| **MnCuY** | ✓ | ✓ | ✓ II ph. | | ↑ | ↓ | | | ↑ | | | | | |
| **MnY** | ✓ | ✓ | ✓ II ph. | | ↑ | ↓ | ↓ | | | | | | | |
| **MnCe** | | ✓ | ✓ II ph. | ✓ | ↑ | ↓ | | | ↓ | | = | = | = | |
| **MnMm** | | ✓ | ✓ II ph. | | ↑ | ↓ | | Sloped | ↑ | | | | | ✓ $O_2$ |